# Attomicroscopy: from Femtosecond to Attosecond Electron Microscopy

**Mohammed Th. Hassan**

Department of Physics, University of Arizona, Tucson, Arizona 85721, USA

Correspondence to: E-mail: mohammedhassan@mail.arizona.edu

**Abstract**

In the last decade, the development of Ultrafast Electron Diffraction (UED) and Microscopy (UEM) has enabled the imaging of atomic motion in real time and space. These pivotal table-top tools opened the door for a vast range of applications in different areas of science spanning chemistry, physics, materials science, and biology. We first discuss the basic principles and recent advancements, including some of the important applications, of both UED and UEM. Then, we discuss the recent advances in the field that have enhanced the spatial and temporal resolutions, where the latter, however, is still limited to a few hundreds of femtoseconds, preventing the imaging of ultrafast dynamics of matter on the scale of several tens of femtoseconds. Then, we present our new optical gating approach for generating an isolated 30 fs electron pulse with sufficient intensity to attain a temporal resolution on the same time scale. This achievement allows, for the first time, imaging the electron dynamics of matter. Finally, we demonstrate the feasibility of the optical gating approach to generate an isolated attosecond electron pulse, utilizing our recently demonstrated optical attosecond laser pulse, which paves the way for establishing the field of "Attomicroscopy", ultimately enabling us to image the electron motion in action.

## 1. Introduction

Time and its definition have excited scientists' curiosity for ages. Although there is no clear definition of time yet, the undeniable fact is that "time" exists. Hence, our world and our life are dynamic, and events occur in a sequence. Taking a static picture of an event does not tell the full story and does not answer this important question: "how does this event happen and evolve in time?" However, a consecutive series of images or a video could provide all the information needed to explain that event. This applies for both macrocosm and microcosm dynamics at different scales and levels of space and time. For instance, in the classic chemistry example, molecules in their initial state before a reaction starts, as well as the new molecular products formed after the reaction finishes, have well-defined structures; however, we do not know how this reaction happens and if there are any transition states. Previously, it was thought that chemical processes were "immeasurably fast"[1]. However, the introduction of femtosecond time-resolved spectroscopy based on the principle of "time freezing" allowed, for the first time, access to the structural dynamics in chemical reactions. This provided more information about the transition states, which allowed control of these reactions and their final products [2]. This revolutionary achievement in science gave birth to ultrafast science and have allowed access to the atomic motion of matter in real time [3]. Moreover, in the last decade, the development of attosecond physics and spectroscopy permitted the real-time observation of electron motion in atoms, molecules, and solid state [4-9]. Although ultrafast spectroscopy provides important information on the transient intermediates of matter dynamics, it does not give sufficient information about the mechanism of this dynamics and its trajectory(ies) in the spatial domain. Instead, one has to rely on theoretical models to gain some insight into the pathways of the atomic and/or electronic motion in the transition states. Hence, envisaging this motion in both domains is highly demanded. Accordingly, making a video of the atomic motion and molecular dynamics is one of scientists' greatest aspirations [10-15]; that is, to watch these motions as they occur, which could provide information on the dynamics in both coordinates to fully understand the natural phenomena of interest. This raised an important question: "what type of "camera" could be used to image the ultrafast atomic motion with high spatial resolution?"

In the last century, the development of electron microscopy and X-ray diffraction provided a remarkable tool for imaging and resolving the three-dimensional structure of matter with atomic resolution, which had a great impact in different fields [16, 17]. Recently, the fourth-dimension "time" has been introduced for probing matter dynamics by utilizing electron bursts. The generation of ultrafast (picosecond and femtosecond) electron pulses enabled the establishment of Ultrafast Electron Diffraction (UED), Ultrafast Electron Microscopy (UEM), and Scanning Ultrafast Electron Microscopy (SUEM). These tools enabled the recording of images of the structural dynamics and atomic motion in real time and space. These tools have found numerous applications in chemistry, physics, biology, and materials science [10, 13-15, 18-20]. Currently, the research activities in the field focus on improving both the spatial and temporal resolutions in order to resolve the ultrafast dynamics of matter with better contrast. In electron diffraction and microscopy experiments, the temporal resolution is defined by the ultrafast electron pulse duration and its synchronization stability with the triggering optical pulses. However, the electron pulse suffers from temporal broadening due to the space-charge effect and energy dispersion during its propagation from the source to the sample, which ruins both the temporal and spatial resolutions. Therefore, many electron-pulse compression techniques have been developed for controlling the space-charge effect and generating ultrashort (bright) electron pulses [14-16]. These techniques enable the confinement of the electron pulse to hundred femtoseconds; however, they suffer from time jittering and the temporal synchronization issues, which limit the temporal resolution in time-resolved electron experiments. Therefore, the ultrafast dynamics measurements that have been carried out so far are on the timescale of picoseconds to several hundreds of femtoseconds [21-25]. Hence, imaging of faster dynamics (i.e., electron dynamics) in matter still remains beyond reach.

Recently, we demonstrated generation of the shortest electron pulse (30 fs) in UEM by the optical gating approach, which breaks the conventional compression limits for an electron pulse and attains electron-dynamics-scale temporal resolution in electron microscopy [26]. In this approach, the generated gated electron pulse duration is limited only by the gating laser pulse, which could be on the attosecond time scale [27]. This approach might eventually lead to the generation of isolated attosecond electron pulses and could open the way for establishing a new "Attomicroscopy" field [26]. It will allow the real-time imaging of electronic motion as theoretically studied in atoms, molecules [28], and condensed matter [29], which could radically change our insight into the workings of the microcosm and could hold the promise for breaking new grounds in a number of fields of science and technology.

This review article provides an overview of the developments in the fields of time-resolved electron diffraction and microscopy. The article is structured as follows. After the introduction in Section I, Sections II–IV discuss the basic principles of UED, UEM, and SUEM, respectively. Then, some of the main applications of these tools that have been reported are discussed in Section V. In addition, Section VI presents the recent advancements in the electron source developments, which enhance both the temporal and spatial resolutions. In Section VII, the photon–electron interaction and the "optical gating" approach for generating the shortest electron pulse (30 fs) in UEM with sufficient intensity to image the ultrafast dynamics of matter are explained. Finally, Section VIII presents on the feasibility of exploiting this approach to generate attosecond electron pulses by utilizing optical attosecond pulses and the establishment of "Attomicroscopy".

2. Ultrafast Electron Diffraction

Since the discovery of X-rays in 1895, diffraction techniques permitted the resolution of three-dimensional (3D) structures of different type of systems, from simple structures such as diatomic molecules (NaCl) to more complex systems such as DNA, proteins, and viruses [30, 31]. Later, the discovery of electron in 1897 paved the way for the establishment of electron diffraction, which allows the determination of gas-phase structure, surface structural analysis, and structural determination of biological systems [31]. Electron diffraction has remarkable advantages over the X-ray diffraction technique due to the electron nature, which we will explain at the end of this section. Although the electron and/or X-ray diffraction are powerful techniques for resolving the static spatial arrangements of atoms, the knowledge of structural dynamics remains necessary for revealing the mechanism of atomic motion. Therefore, introducing the time domain in these techniques was necessary.

The earliest efforts on introducing time resolution into electron diffraction were made for short time domains (milli-to-micro second time scale) [20]. Rood & Milledge [32] conducted studies on the radical dynamics in the gas phase by electron diffraction on the sub-millisecond time scale. Then, Bartell & Dibble [33] studied the phase change in clusters that are produced in supersonic jets, with a time-of-flight resolution of 1 μs. Later, Ewbank et al. [34, 35] enhanced the temporal resolution to sub-nanoseconds by combining an intense-laser-initiated electron source with a linear diode array detector. In 1982, Mourou & Williamson [36] introduced the methodology of the modified Bradley–Sibbett streak camera to record the diffraction from thin aluminum films in the transmission mode with 100 ps pulses. Later, 20 ps electron pulses were produced to study the films before and after irradiation with a laser pulse [36]. In the beginning of the millennium, the groups of Zewail [21, 37, 38], Cao [39], and Miller [22] have focused their efforts on the generation of ultrafast electron pulses, which helped attain a few-picosecond and femtosecond temporal resolution, and in establishing UED, which provided real-time access to the atomic motion and enabled the recording of molecular movies.

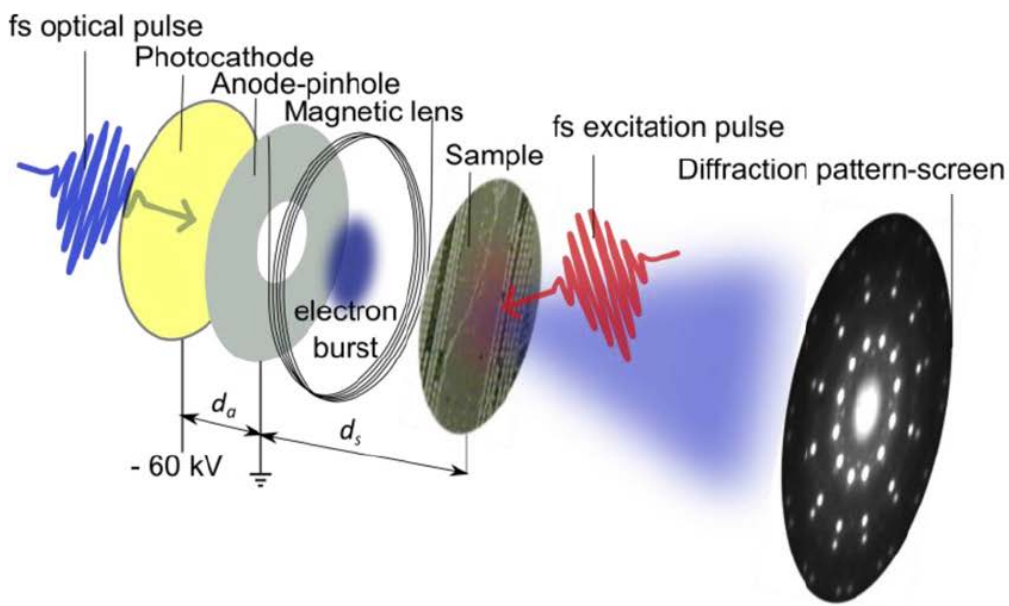

Figure 1. Typical layout of UED. The UV optical pulse is focused on the photocathode to generate a femtosecond electron pulse to probe the sample dynamic, which is pumped by another laser pulse. Then, the electron diffraction pattern is recorded by the detector [18].

The first demonstration and imaging of the transient structure in complex chemical reactions utilizing 1 ps electron pulses has been reported by Ihee et al. [21]. Later, Siwick et al. [22] broke the picosecond temporal resolution limits in UED and demonstrated a fully-resolved structural transition using 600 fs electron pulses.

The UED experiment, with its typical layout shown in Fig. 1, is based on triggering the ultrafast dynamics by a laser pulse and recording the electron diffraction pattern by a CCD camera. The electron diffraction pattern is formed due to the interaction between the ultrafast electron "probe" pulse and the sample under study. A video of the atomic motion can be obtained by acquiring a sequence of diffraction pattern snapshots for different time delays between the trigger and the probe pulses. Then, the temporal dynamics of the 3D structure can be retrieved from the acquired diffraction pattern. Accordingly, the recorded movie provides access not only to the atomic motion dynamics but also to its trajectory and transition mechanism from the ground to the final state, which is beyond the reach of the time-resolved spectroscopy measurements.

In typical UED measurements, the electron source should have extremely high spatial and temporal resolutions, as well as a sufficient intensity to visualize the atomic motions with high contrast. Moreover, the precise determination of the time axis and especially the reference time is essential in this experiment. Therefore, it is crucial to accurately synchronize both the excitation pulse, which alters the dynamics, and the probe electron pulse. Hence, these pulses should be derived from the same laser oscillator.

A key challenge in resolving the 3D structural dynamics from electron diffraction patterns is that the diffraction patterns contain contributions from incoherent atomic scattering as well as the coherent molecular interferences arising from atom–atom pairs, because the electrons scatter off all the atoms and atom–atom pairs in the sample. Therefore, the increase in the signal due to the change in dynamics is very small compared to the background signal. This challenge can be overcome by using the "frame-reference" method. In this method, the dynamic transient can be obtained based on the difference between the recorded diffraction patterns at different delay times and an in-situ reference pattern (which represents the ground-state structure) obtained at a negative time before the arrival of the trigger pulse. One of the important advantages of this methodology, in addition to the isolation and enhancement of the transient dynamics from a small signal change, is the elimination or minimization of the systematic error associated with the detection system [20].

Another technical challenge in UED is the sample preparation. The thickness of the sample under study must be on the order of a few tens of nanometers to ensure that the diffraction process occurs in the single electron scattering limit for simple inversion of the diffraction pattern. In addition, conducting a single-shot time-resolved electron diffraction measurement for irreversible sample dynamics is difficult due to the small number of electrons, which is limited by the space-charge effects. This effect is caused by the inherent electron properties as charged particle (electron–electron repulsion) and leads to broadening and reduction in the density of electron pulses. Recently, electron-pulse compression techniques have been developed to overcome the space-charge effect and limit the electron pulse duration in order to increase the electron density and enhance the temporal resolution.

Further development of UED led to the establishment of Ultrafast Electron Crystallography (UEC) to study the ultrafast dynamics of crystalline solid-state systems [20, 40, 41] (the UEC setup developed by the Zewail group is shown in Fig. 2). There are two modes for UEC (Fig. 2a): transmission and reflection modes. In the transmission mode, which provides a better temporal resolution, the elastic mean free path for nonrelativistic electrons is on the order of tens of nanometers and ~ 5 times longer when MeV electrons are used [42-44]. This fact generally constrains the sample thickness to be on the order of a few tens of nanometers and makes the sample preparation one of the biggest technical challenges in the transmission electron diffraction studies [45]. In addition, the sample and the host grade must be perfectly flat to minimize the temporal jittering across the sample surface at the reference time. The optimum alignment of the sample tilt can be performed using optical interference. The ultrathin sample allows the achievement of the perfect match between the optical excitation depth and electron probing, which ensures a homogenous excitation level across the sample thickness [18].

On the other hand, the reflection mode is utilized to study the surface structural dynamics of bulk samples [40, 41, 46-52]. In this arrangement, the incident angle of the electron beam is a few degrees, while the pump laser pulse comes almost perpendicularly to the sample surface. The sample is placed on a controlled stage to allow for the relative orientation control between the surface normal and the incident electron beam direction. This technique has been used to study the phase transition in a single crystalline $VO_2$ sample, utilizing the optically tilted wave-front to reach resolution on the sub-ps level [53, 54] (Fig. 2(c)). In the time-resolved electron diffraction measurement utilizing this reflection mode, one should be very careful due to the surface sensitivity. Therefore, a relatively high excitation fluence of the pump laser pulse is required to induce a structural change signal that is sufficiently above the noise level. However, this could generate surface dipoles via Multiphoton Photoemission (MPPE), Transient Electric Fields (TEF), and an excess of positive charges at the surface [55-58], which affects the measured diffraction pattern and causes misinterpretation of the experimental results. Hence, this should be taken into account when performing the experiment and data analysis.

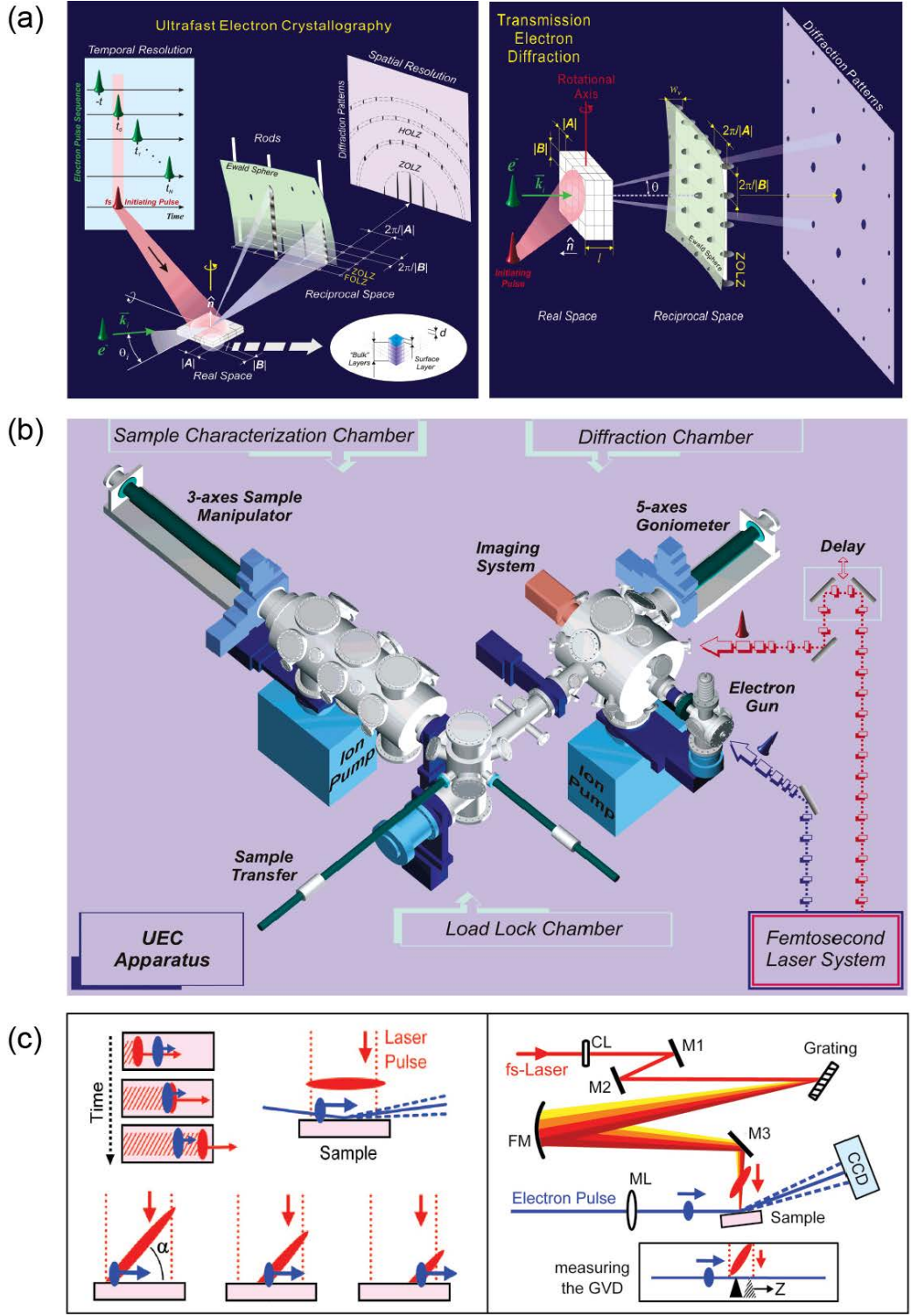

Figure 2. Ultrafast Electron Crystallography (UEC). (a) Schematics of the UEC arrangements in reflection and transmitted modes (40). (b) Illustration of the UEC setup developed by Zewail group [40]. (c) The scheme of the optically tilted wave-front approach to overcome the phase mismatch and achieve the maximum temporal resolution in UEC [54].

Alternatively, time-resolved X-ray diffraction techniques have been developed for resolving the structural dynamics in real time with high spatial resolution [13]. However, UED has the following advantages over the X-ray techniques [59-66]. First, the cross-section for electron scattering is ~ 6 orders of magnitude larger than that for X-ray scattering [10, 67]. Second, the quantum efficiencies for generating electron pulses based on photoemission are on the order of $10^{-4}$ and $10^{-1}$ for metals and semiconductor photocathodes, respectively [68], whereas the conversion efficiency for the generation of X-ray photons is typically $10^{-6}$ [69-71]. This efficiency is further reduced due to the limited solid angle of collection and X-ray optics by a factor of ~ $10^{-2}$ at the position of the sample. Third, the electron–matter interaction is stronger, and therefore the electron diffraction can reveal the transient structure of gases, surfaces, and (thin) crystals. In addition, damaging to specimens per useful elastic scattering event is lower for electrons [72].

Finally, the development of UED enabled the study of transient structural dynamics with respect to time and space (i.e., the non-concerted elimination reaction of dihaloethanes) in addition to the study of the excited-state structures in chemical reactions. Accordingly, the implementation of UEC opened the door for many applications to study the solid-state phase transition and surface dynamics. In Section V, we will explain some of these applications in detail. Moreover, the advances in UED and UEC led to the establishment of UEM, which adopts the same basic principle, that is, the generation of ultrashort electron pulses inside the Transmission Electron Microscope (TEM). UEM allows connecting the ultrafast dynamics of matter with its morphology, direct imaging of the structural changes, and ultrafast electron spectroscopy, which will be discussed next.

3. Ultrafast Electron Microscopy

TEM is one of the most powerful known imaging instruments [16]. The first TEM was invented in 1933 by Max Knoll and Ernst Ruska at the Technical College in Berlin. In TEM, a beam of electrons passes through a thin sample followed by a series of lenses, forming a highly magnified image of the sample on a screen. In contrast to the optical microscope—for which the spatial resolution is limited by the wavelength of light—the TEM allows the resolution and imaging of 3D structures on the atomic scale, which has a significant impact in different fields of science. However, the temporal resolution of TEM is limited by the video-camera recording rate (millisecond), since the electrons are produced by heating the source or by field emission, resulting in randomly distributed bursts of continuous electron beams. For breaking this millisecond temporal resolution limit in TEM to image fast dynamics events, Bostanjoglo et al. pioneered the use of pulsed capacitors to produce electron beam flashes [73-77]. This helped in achieving nanosecond resolution, which has been used to study the melting process of a 50 nm amorphous NiP film that was exposed to a 7 ns laser pulse [76]. Later, both the temporal and spatial resolutions were improved at LLNL-Berkeley [78-80] to perform time-resolved experiments following the same approach. Then, Kim et al. [79] demonstrated the first single-shot imaging with nanometer resolution by using a high-brightness 15 ns electron pulse. Recently, the Four-Dimension Ultrafast Electron Microscope (4D UEM) was developed by Zewail's group [14, 81], in which picosecond- and femtosecond-scale temporal resolution—10 orders of magnitude better than that of conventional electron microscopes—was achieved. This became possible by generating ultrafast electron pulses via photoelectron emission from the microscope cathode. In this case, the temporal resolution becomes independent of the rate of camera recording and is limited only by the electron pulse duration. Different techniques, in addition to direct imaging, have been demonstrated utilizing the UEM, such as electron diffraction [81], Photon-Induced Near-Field Electron Microscopy (PINEM) [82], and Electron Energy Loss Spectroscopy (EELS) [83].

In UEM, the high-contrast imaging requires a large number of electrons. However, due to electron repulsion, the large number of electrons in the pulse wave packet imposes limits on the temporal and spatial resolutions of the microscope. Therefore, the temporal resolution can be maintained with single-electron imaging, where coherent and timed single-electron packets can provide an image equivalent to that obtained by using many electrons in conventional microscopes. The electron wave packet—confined into a femtosecond time window—has a unique coherence volume, which depends on the electron velocity. The image emerges at certain instances of time when a sufficient number of electrons click the detector.

For the electron beam generated in a UEM, the temporal (longitudinal) coherence length of the electron packet is given by $L_{tcl} = v_e(h/\Delta E)$, where $\Delta E$ is the energy of the photoelectrons relative to the work function of the cathode, $v_e$ is the electron velocity, and $h$ is Planck's constant.

The spatial (transverse) coherence length (electron beam coherence) is determined by $L_{scl} = (h/\Delta P_T)$, where $P_T$ is the transverse momentum spread. In addition, $L_{scl}$ can be expressed in terms of the source angular deviation $\alpha$ at the specimen as $L_{scl} = (\lambda_e/\alpha)$, where $\lambda_e$ is the de Broglie wavelength. Moreover, $\alpha = d/L$, where d is the source width, and L is the distance to the specimen from the source. Therefore, $L_{scl}$ can be calculated from $L_{scl} = (\lambda_e/\alpha)$. By determining both $L_{scl}$ and $L_{tcl}$, the coherence volume $V_c\ (cell)$ of each cell can be obtained from

$$V_c\ (cell) = \Delta x \Delta y \Delta z = L_{tcl}(longitudinal) * L_{scl}(transverse),$$

i.e., the coherence volume $V_c$ of the electrons accelerated to 200 keV inside the UEM (de Broglie wavelength $\lambda_e = 2.5079$ pm) is $10^6$ nm$^3$. The coherent cell volume and the source brightness are important parameters for defining the imaging resolution in the electron microscope [14].

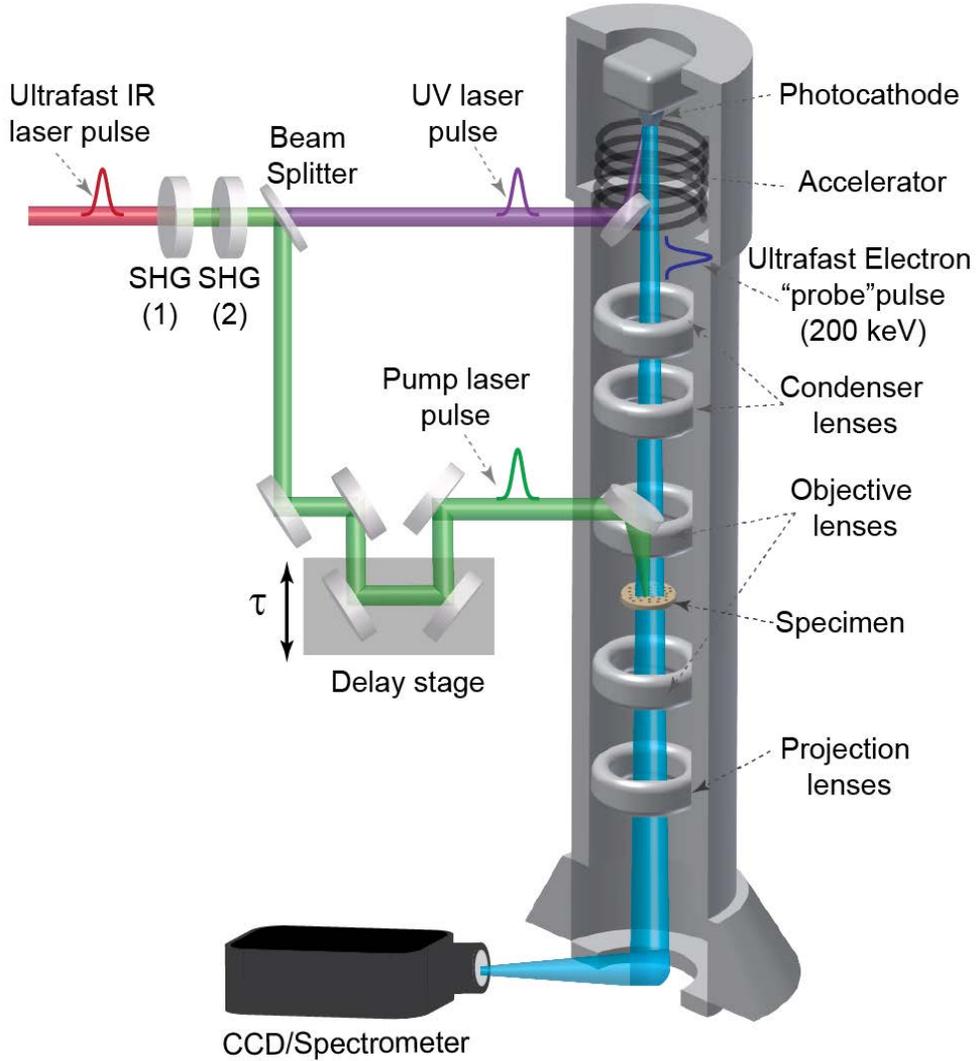

Figure 3. Illustration of the time-resolved UEM experiment. Two laser pulses (DUV and Visible) are generated from the same laser source by harmonic generation processes. The UV pulses are directed to the photocathode inside the microscope to generate femtosecond electron pulses, which are accelerated and focused on the sample under study. The visible laser pulse alters the system dynamics, which can be probed by different modes of the microscope (direct imaging, electron diffraction, and Electron Energy-Loss Spectroscopy (EELS).

In the UEM time-resolved measurements, the ultrafast electron burst acts as a "probe" pulse—similar to the camera shutter—as such, this "freezes" the motion of the atom, with the corresponding photographs forming a real-time movie of the ultrafast process triggered by the laser "pump" pulse. The time axis is defined by the relative delay between the electron probe pulse and the laser pump pulse. The latter defines the reference time point (time zero) for the evolution of the atom motion.

In the UEM experimental setup, which is illustrated in Fig. 3, Deep Ultraviolet (DUV) laser pulses are focused by an external lens on the photoemissive cathode inside the microscope to generate ultrafast electron pulses via the photoemission process, and then the generated electrons are accelerated (200 keV) in the microscope column, which is maintained under high vacuum. Then, the electrons pass through the condenser lens system, which controls and reduces the diameter of this beam. It consists of two lenses; the first one (strong lens) demagnifies the image of the electron source to provide a small point source for increased coherence. The second lens is weaker, and it projects the demagnified source image onto the specimen. This lens controls the illumination spread on the screen. Underneath these two lenses is the condenser aperture, which collimates the electron beam and modifies its intensity. The specimen is held in place by a sample holder within the field of the objective lens, which is located just on top of the sample chamber. The tilt of the sample for certain types of imaging measurements can be controlled automatically. After the sample, there is another objective lens and an aperture. The objective lens aperture filters out the beams that come out of the specimen in a particular range of angles. The selected area aperture filters out the beams that come from a particular set of positions in the specimen. The plane to map the beams transmitted from the sample at the detector is chosen based on the measurement mode (diffraction mode or image mode) by changing the excitation of the projector lens. Hence, either an image or a diffraction pattern can be obtained. The microscope can be equipped with an energy analyzer and electron spectrometer, which enables electron spectroscopy experiments. A laser pulse can be directed to the sample by some optical elements (mirrors and a lens). This pulse can pump the system, which is then imaged by the electron pulse. In the time-resolved electron microscopy measurements, the delay between the electron and laser pulses is controlled precisely by a linear delay stage.

The different modes of UEM (direct imaging, diffraction, and electron spectroscopy) allow different applications in imaging the ultrafast dynamics in the solid phase. Moreover, we recently demonstrated the first ultrafast electron microscopy in an aqueous solution by implementing a liquid cell inside the UEM [84]. This work opens the door to the study and imaging of the ultrafast dynamics of chemical reactions and biological systems in their native environments. Some of these applications will be discussed briefly in the corresponding section.

## 4. Scanning Ultrafast Electron Microscopy

Scanning Ultrafast Electron Microscopy (SUEM) is based on the same idea as UEM. However, the electron source in SUEM is a field emitter with a tip of tens to hundreds of nanometers in size, which provides an electron beam of higher brightness than the source in UEM, where the photocathode has an active area with a size of tens of micrometers. The SUEM instrument is used mainly for imaging surface dynamics, so that the sample preparation is easier (ultrathin samples are not required). The first efforts to introduce the time dimension in Scanning Electron Microscope (SEM) were made by "chopping" the electron beam using the high-frequency (MHz or GHz) beam deflection and blanking technique [85, 86]. The nanosecond temporal resolution, obtained by this technique, has been utilized to study the response of microelectronic devices and the vibrational mechanics of microstructures under the influence of variable voltages [87]. The enhancement of the spatiotemporal resolution to the picosecond scale enables the study of the carrier dynamics in cathodoluminescence, which is induced by the electrons and detected by a streak camera [88, 89].

The femtosecond resolution has been obtained by the generation of ultrafast electron pulses based on the photoemission process at the tip in SUEM. In this case, the temporal resolution is defined by the temporal width of the electron pulse and not by the deflection rate [86] or the parameters of the streak camera used for optical detection in [88]. Similar to that in the case of UEM, the electron pulse acts as a "probe" pulse to image the surface dynamics pumped by a laser pulse and does not react with the sample under study, as in the case of the study reported elsewhere [88].

Although SUEM adopted the same pump–probe scheme as used in UEM, the detection mechanism is completely different, since it is based on pixel-by-pixel recording rather than parallel processing of the image, as in the case of UEM. Moreover, the observed signal in SUEM is obtained in the form of secondary or backscattered electrons, which mainly result from inelastic scattering. In addition, the Electron Backscattering Diffraction (EBSD) patterns recorded in SUEM consist of Kikuchi lines [90, 91] instead of the Bragg spots and Debye–Scherrer rings in the recorded diffraction patterns in UEM. Therefore, SUEM is utilized to study the evolution of structural dynamics in crystalline grains or domains along different crystallographic directions [92, 93].

The SUEM experiment setup developed by the Zewail group is shown in Fig. 4 [92, 93]. In this experiment, a high-repetition-rate (tens of MHz) and high-power laser system is required to generate powerful ultrashort laser pulses. Through a nonlinear process, ultrashort (few hundred femtosecond) UV pulses—with photon energies higher than the work function of the tip—are generated from the original laser beam before it is tightly focused (the average energy on the order of several nJ) onto the sharp tip of the emitter-electron inside the SUEM to generate the femtosecond electron "probe" pulse, which is accelerated to a few tens of keV before it illuminates the sample. Another laser pulse, which emerges from the same laser source to ensure perfect synchronization and perfect time referencing, is directed to the sample to trigger the ultrafast structural dynamics of the system under study. The delay times (time-axis) between the pump and probe pulses are precisely controlled with a delay stage.

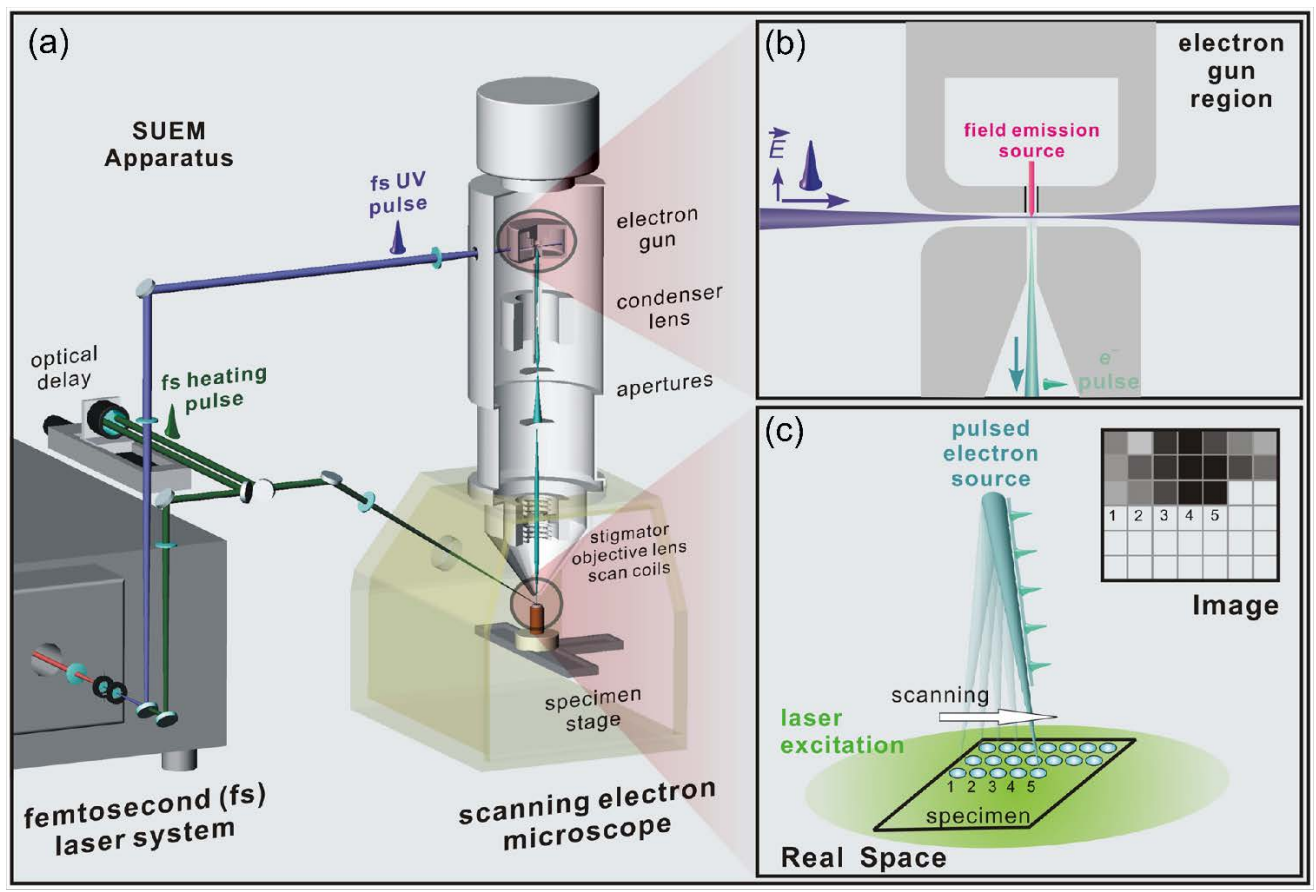

Figure 4. Scanning Ultrafast Electron Microscopy (SUEM). (a) Illustration of the conventional SUEM setup, modified from typical SEM. A close-up view of the field-emission and the generation of ultrafast electron pulse is shown in (b), while (c) shows the pixel-by-pixel image construction in SUEM [92].

In the acquisition mode, at a certain time delay between the laser "pump" or "trigger" and electron "probe" pulses, the focused pulsed electron beam is directed by the scan coils and raster scans across the specified region of the specimen to form an image (Fig. 4). The scanning across the specimen, with perfect timing, is performed to record a constructive image based on mapping over discrete pixels in space. The electron–matter interaction results in various types of signals, such as secondary and backscattered electrons, as well as X-rays, and these are recorded by different detectors [90, 94]. A full recorded movie of the structural dynamics can be obtained by repeating the spatial scanning and constructing pixel-by-pixel images for different delay times. Recently, some interesting applications that utilize SUEM have been reported to directly image the carrier dynamics in solid-state. This will be discussed in the next section.

**5. Applications: from chemistry to biology**

Over the last two decades, UED, UEC, UEM, and SUEM have found a vast range of applications spanning chemistry, physics, biology, and materials sciences. In this section, we will discuss some of these applications in detail.

*5.1. Chemistry: ultrafast molecular structural dynamics*

In the first demonstration of UED by Ihee et al. [21], the ultrafast dynamics of the chemical reactions of 1,2-diiodotetrafluoroethane ($C_2F_4I_2$) to produce tetrafluoroethene and iodine, as well as the ring opening of 1,3-cyclohexadiene (CHD) to form 1,3,5-hexatriene, were studied with a spatiotemporal resolution on the order of 0.01 Å and 1 ps. The first reaction occurs in two steps, as seen in the temporal evolution diffraction measurements (Fig. 5(a)). The first step ($C_2F_4I_2 \rightarrow C_2F_4I + I$) is completed within the first 5 ps of the reaction. Then, the second step ($C_2F_4I \rightarrow C_2F_4 + I$) takes place on a scale of 31 ± 4 ps. Unlike this reaction, the second reaction (ring opening of 1,3-cyclohexadiene (CHD)) involves structural rearrangement rather than fragmentation, which also takes place on a picosecond time scale, as shown by the electron diffraction measurements (Fig. 5(a)). In addition, UED has been used to study the radiationless transitions of four prototypical hetero-aromatic (pyridine, 2-methylpyridine, and 2,6-dimethylpyridine) and aromatic-carbonyl (benzaldehyde) organic molecules as reported in [95]. In this study, the electron diffraction pattern was recorded for the initial states. Then, the change in this diffraction pattern was traced in time. The retrieved ultrafast dynamics showed that the parent structure had essential influence on the dynamical evolution, relaxation pathways, and their respective time scales.

Later, the development of UEC allowed the study of structural dynamics of interfacial water following substrate photo-induced heating [41]. This study gave more insights about the transformation dynamics from an ordered to a disordered structure. Their coexistence depends on the time scales of the movements of atoms, both locally and at a long range of interfacial water. In this work, the invoked probe electron pulses had a de Broglie wavelength of λ = 0.07 Å at 30 keV and a temporal duration of a few hundred femtoseconds. The interfacial water was formed on a hydrophobic surface (silicon, hydrogen-terminated). Hence, the interfacial and ordered (crystalline) structures were extracted from the Bragg electron diffraction, while the layered and disordered (polycrystalline) structures were obtained from the Debye–Scherrer rings.

First, only the electron diffraction pattern of the Si substrate was recorded; this pattern consists of Bragg spots. Then, water was added onto the substrate surface, and interfacial ice was formed at 110 K. The Bragg spots were transformed into new spots and rings, characteristic of water. The substrate was locally heated by an infrared (IR) pulse, and the electron diffraction pattern was recorded as a function of the time delay between the heating (pump) IR and electron (probe) pulses. The results demonstrated that the restructuring dynamics of the time-dependent long-range order of interfacial water is slower than the time for amorphization, a process where the O···O correlation is lost before the OH··O correlation, and the time scale for losing the hydrogen bond network is on the order of 37 ps.

Recently, major advancements in the generation and compression of ultrafast electron pulses by the RF compression techniques improved the temporal resolution to several hundred femtoseconds [96, 97] and increased the brightness of the electron beam for probing the ultrafast dynamics in more complex systems. For example, Gao et al. [24] took advantage of these short compressed electron pulses to study the phase transition in an organic material (organic salt (EDO-TTF)$_2$PF$_6$), which has a weak scattering center. The photo-induced insulator-to-metal phase transition of that organic salt was investigated by conducting single-shot electron diffraction measurements. The structural dynamics was mapped by recording and tracing the changes in the Bragg reflections. The results (Fig. 5(b)) show that the system undergoes the phase transition in two steps. In the first step, a transient intermediate structure is generated in the early stage of charge delocalization (< 5 ps); in the second step, the system is converted into a metallic structure in hundreds of picoseconds. This high-brightness source can also be used to study the ultrafast photo-induced charge transfer in a more complex system as reported in [98].

The generation of MeV ultrafast electron pulses in UED at SLAC, as explained earlier, achieves significant enhancement of the signal-to-noise ratio with shorter acquisition times and provides a powerful camera for recording molecular videos in real space and time with a very high resolution. Yang et al. [99] utilized this powerful tool to image the photoexcited coherent motion of a vibrational wave packet on iodine molecules (I$_2$) in the gas phase. The images of this motion were recorded with a sub-angstrom resolution in space and 230 fs resolution in time. In this experiment, after laser excitation, the MeV electron diffraction patterns of the system were recorded at different delay times (Fig. 5(c)). The authors demonstrated that, with this resolution, the diffraction pattern became sensitive to both the position and shape of the nuclear wave packet.

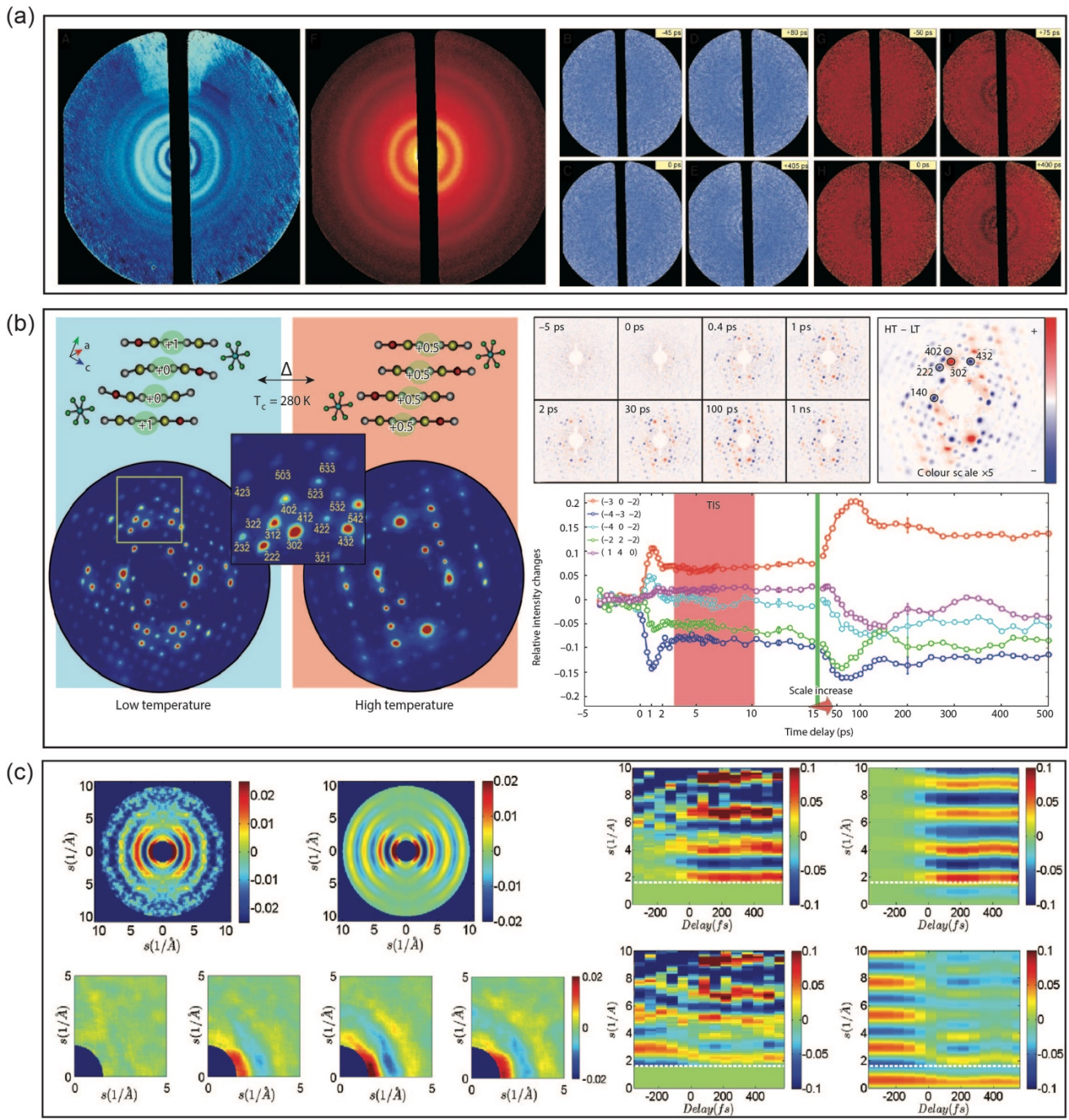

Figure 5. Imaging the Ultrafast Structure Dynamics in real time and space. (a) The electron diffraction pattern of 1, 2-diiodotetrafluoroethane ($C_2F_4I_2$) and 1, 3-cyclohexadiene (CHD) at different delay times [21]. (b) The ultrafast phase-transition dynamics of the organic salt— (EDO-TTF)$_2$ PF$_6$— from insulator to metal state [24]. (d) The MeV femtosecond electron diffraction study of the iodine molecular dissociation [99].

*5.2. Valence and core-level electronic dynamics*

Electron Energy-Loss Spectroscopy (EELS) is a prevailing technique in UEM, which is applied for probing the electron distribution in both valence and core-level excitation with nanometer-femtosecond spatiotemporal resolution. One of the examples of probing the valence electron excitation is demonstrated by Carbone et al. [83]. The authors studied the chemical bonding dynamics in graphite, which was excited by a femtosecond laser pulse, by tracing the change in the Electron Energy-Loss (EEL) spectrum in the low-energy region (< 50 eV). This energy range carries information about the plasmonic properties of the bonding carrier, which reveal the collective dynamics of the valence electrons. In this work, the time-resolved EELS results reflect the change in the electronic structure (sp2/sp3-type hybridization), which causes contraction towards the diamond structure and expansion towards the graphene structure in the graphite atomic planes. On the other hand, time-resolved EELS allows probing of the core-level electron excitation dynamics by acquiring data in the high-energy region (> 100 eV) of the EEL spectrum. Hence, this technique has been utilized in [100] to the study photo-excitation of graphite. The authors probed the carbon K-edge spectrum in the energy range of ∼ 280–450 eV, following ns and fs optical excitations. The results showed a local increase in the C–C bond length within the *ab*-planes of graphite, although the overall behavior of the lattice was still represented by in-plane contraction. These results highlight the important role played by the anharmonicity of the interatomic potential, which is responsible for the increase in the local C–C bond length, while the out-of-plane phonons are responsible for the long-range in-plane contraction in the overall lattice dynamics of graphene and graphite [101].

*5.3. Materials science: ultrafast phase-transition dynamics in solid-state*

Breaking the limits of picosecond temporal resolution in electron diffraction measurements was first reported by Siwick et al. [22], which opened the door for imaging the atomic motion in solid-state phase-transition processes. Siwick et al. [22] studied the photo-induced solid–liquid phase transition (melting) process in aluminum. In this study, the long-order change was traced as a function of time. The 20-nm-thick Al sample was excited by a laser pulse with fluence of 70 mJ/cm$^2$. The long-range order (present in the crystalline phase) disappears, and a short-range atomic correlation (present in the liquid phase) emerges, which is a clear indication of a complete phase transition process occurring within 3.5 ps (Fig. 6(a)). On the other hand, the high-brightness electron beam generated by the RF compression technique has been used to study the structural evolution associated with melting of 20-nm-thick free-standing 111-oriented polycrystalline gold films [23]. This structural dynamic process occurred due to the high-level excitation (14 times the energy required to melt the sample starting from room temperature) using intense UV laser pulses (470 J/m$^2$). Owing to this high-level excitation, the gold nanofilm undergoes a phase transition from solid to high-density plasma, which is also referred to as warm dense matter. From the diffraction dynamics measurements, the authors concluded that the lattice was rapidly superheated and melted homogeneously on the time scale of ~ 1 ps, and the rise time of the liquid peak in gold under the above-mentioned excitation conditions was 7 ps, delayed by 1.4 ± 0.3 ps. The appearance of the liquid structure signature reflects the transition from an anisotropic polycrystalline state into a fully isotropic disordered state (Fig. 6(b)). This experiment was also conducted at a higher excitation level, which led to the acceleration of the dynamics, where the retardation time of the liquid structure signature dropped to be on the order of a few hundred femtoseconds.

Accordingly, as explained in connection with the previously described applications, the melting phase transition of gold is based on the transfer of thermal energy between the excited electrons and the initially cold lattice; however, in other materials (such as semiconductors), the melting phase transition is non-thermal and is caused by the anti-bonding character of the conduction band, which results in lattice collapse. Sciaini et al. [102] studied the structural changes in crystalline bismuth as it undergoes a phase transition (melting) due to laser excitation. The structural dynamics was probed with time-resolved femtosecond electron diffraction. From the measurements, the time scale of this photo-induced melting was resolved and found to be on the sub-vibrational time scale (190 fs). The authors attributed this fast melting transition to a change in the potential energy at the surface of the lattice, which causes strong acceleration of the atoms along the longitudinal direction of the lattice and efficient coupling of this motion to an unstable transverse vibrational mode. The obtained diffraction dynamics curves (illustrated in Fig. 6(c)) exhibit decaying and rising characters for the crystalline and liquid-phase Bi, respectively.

Furthermore, the improvement of the temporal resolution in UED to a few hundred femtoseconds (< 250 fs) allowed the resolution of the atomic motion in the structural dynamics of quasi-two-dimensional Charge-Density-Wave (CDW) materials, as demonstrated in [103]. In this experiment, the ultrafast optically induced phase transition of a 1T-TaS$_2$ system from the Periodic Lattice Distortion (PLD) phase to CDW phase due to the change in the electronic spatial distribution was investigated by conducting time-resolved electron diffraction measurements. The time evolution of the relative change in the Bragg peak diffraction signal, triggered by the photoexcitation, indicated that the intensity of the CDW diffraction signal is suppressed by 30% on the time scale of hundreds of femtoseconds; simultaneously, a related increase occurs in the Bragg scattering intensity due to the optically induced redistribution of the electron density (Fig. 6(d)). This work demonstrated the ability of UED to directly observe atomic motion on time scales that are short enough to even follow the effect of non-equilibrium electronic distributions on strongly correlated lattice dynamics, as recently demonstrated in [104, 105].

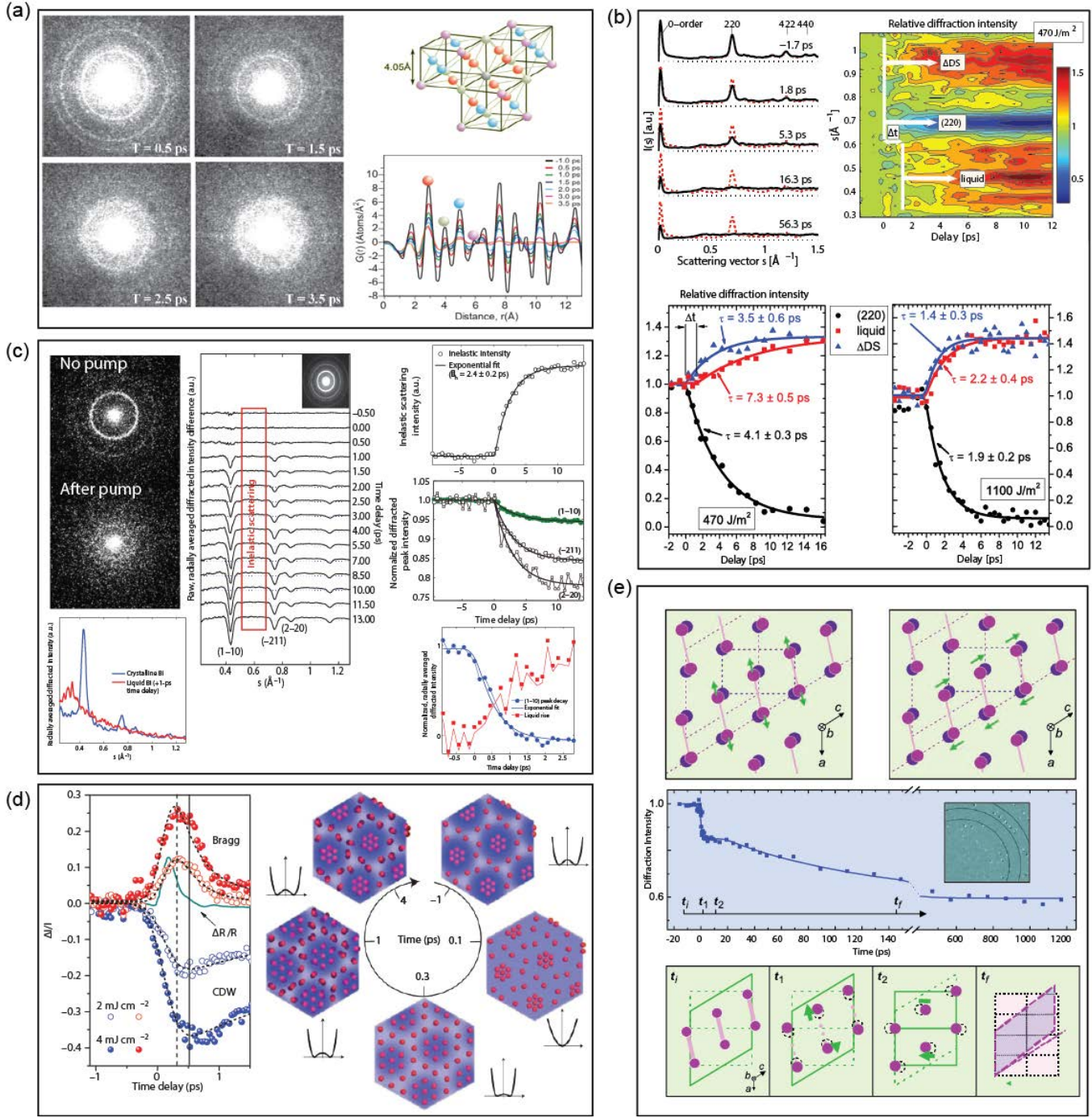

Figure 6. Ultrafast phase-transition dynamics in solid-state. (a) The electron diffraction measurements for the photo-induced phase transition (melting) of Al utilizing femtosecond electron diffraction [22]. (b) The study of ultrafast phase transition of polycrystalline gold nanofilm with a single shot UED technique [23]. (c) The electron diffraction and dynamics curves for the structural change of the crystalline bismuth melting process [102]. (d) The phase transition of 1T-TaS$_2$ system from periodic lattice distortion (PLD) phase into CDW phase [103]. (e) The temporal evolution of the vanadium atomic motion in the ultrafast phase transition process of bulk VO$_2$, which is studied using UEC in [53].

Alternatively, the critical development of UEC aimed to overcome the phase-matching problem by the implementation of the "tilted geometry" illustrated in Fig. 2(c) enabled the necessary atomic-scale spatial and temporal resolutions to study phase transitions and related atomic motion in solid-state systems in real time and space. Gedik et al. [49] reported the non-equilibrium phase-transition dynamics in cuprates (oxygen-doped $La_2CuO_{4+\delta}$), which has superconductivity properties below the critical temperature (Tc) and metallic properties at room temperature. In this study, the static electron diffraction was recorded first, and the lattice constant was defined from the Bragg spots. Then, the diffraction patterns as functions of the arrival time of the pump laser pulse, relative to the probe electron pulse, were acquired, and the change of the lattice constant was retrieved. The observed changes were on three different time scales: 5 ps and 27 ps, attributed to the formation of the transient phase, and 307 ps for structural recovery. The results also proved that the phase transition dynamics depend on the fluence of the trigger laser pulse.

Another important application of UEC, which utilizes the reflection mode, is the study of the phase transition in bulk vanadium dioxide ($VO_2$) from the insulator (monoclinic) to metal (tetragonal) phase demonstrated in the study of Baum et al. [53]. The phase-transition dynamics were extracted from time-resolved electron diffraction measurements. In this experiment, the bonding electrons of the vanadium pairs, which are responsible for the stabilization of the initial monoclinic structure, were photo-excited by an ultrashort laser pulse. The excitation of the anti-bonding states results in a repulsive force, which moves the vanadium atoms within the V–V pairs apart from one another on a time scale of several hundred femtoseconds. The long-range displacement of atoms within the unit cell, due to electron–phonon coupling, takes place on a longer (picosecond) time scale. Then, on a time scale of hundred picoseconds, the acoustic shear waves drive the lattice toward the final tetragonal structure (Fig. 6(e)). These observations indicate that stepwise atomic motion, rather than direct structural conversion, mediates the phase transition in $VO_2$. Alternatively, in a study by Hassan et al. [106], the dielectric response of the $VO_2$ nanoparticle crystalline structure during the insulator-to-metal phase transition was probed utilizing a different technique, namely the time-resolved PINEM. From the measured dynamics of the PINEM intensity change and comparison with ultrafast diffraction data, the authors were able to retrieve the change in the dielectric constant associated with the stepwise atomic motion of the transition, opening up the possibility of studying phase transformations from the electronic, rather than the structural, perspective. This technique will be explained in detail later. Morrison et al. [25] studied the same system, but in the polycrystalline form, focusing on the exploration of the electron–lattice interaction and electron–electron interaction dynamics. In that study, the authors used a combination of UED and time-resolved IR transmittance measurements, which provided more insights on the phase-transition process. The results demonstrated a photo-induced transformation into a metastable state that not only retained the periodic lattice distortion characteristics of the semiconductor but also acquired metal-like mid-infrared optical properties.

*5.4. Imaging charge carrier dynamics in real time*

SUEM, with ultrashort temporal and nanometer spatial resolutions, was exploited to directly image the carrier interface dynamics in p–n junctions excited by laser pulses as demonstrated in the study by Najafi et al. [107]. In this work, the electron pulse (accelerated at 30 kV) was generated from a field emission tip by illumination with UV laser pulses and focused on the surface of the sample. Then, the signal was collected by a secondary electron (SE) detector (Fig. 7). The image, in the absence of the pump laser, was obtained by spatially scanning (with 200 nm step size) the sample surface. In the acquired image, the difference in brightness between the p-type and n-type regions was clearly observed, which was due to the difference in effective electrons between the two regions (Fig. 7(a)).

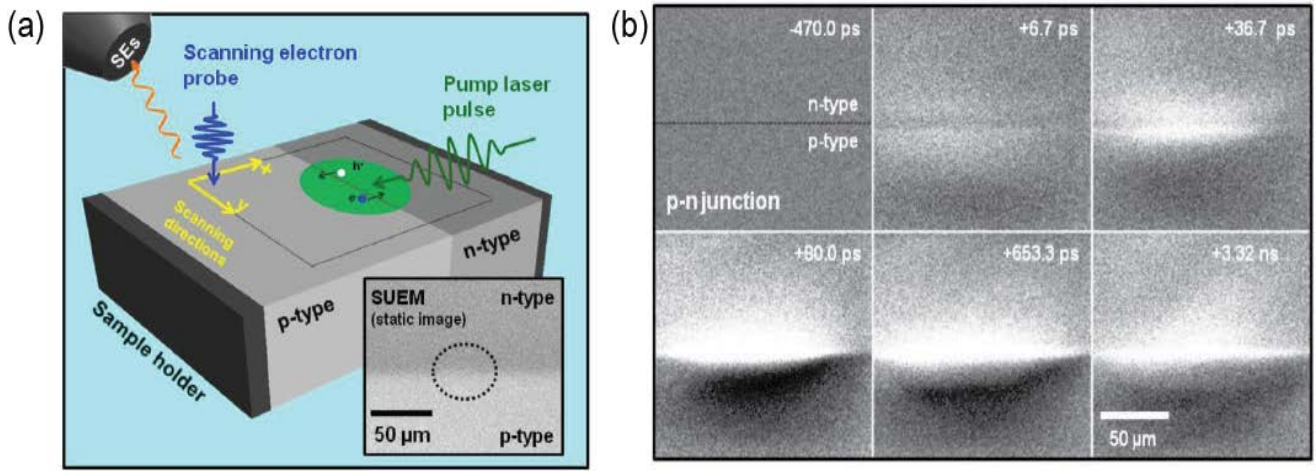

Figure 7. Imaging the charge carrier dynamics in real time and space by SUEM. (a) Illustration of the cross surface acquiring image of the charge carrier dynamics in the p-n junction. (b) Snapshots of the charge carrier diffusion at different time delays [107].

The visible laser pulse then excited the sample, which altered the carrier excitation, with diffusion to the conduction band. Snapshots were recorded at different instants of time (Fig. 7(b)) with 2 ps time steps. The bright contrast between the p and n regions gradually decays with time. From the images shown in Fig. 7(b) we can observe that both layers are bright at + 6.7 ps, indicating their individual behavior in the absence of a junction. However, at 36.7 ps, the charge carriers start moving towards the junction, resulting in excess electron and hole densities of the n-type and p-type, respectively. The depletion layer at the junction remains dark due to the surface patch fields that hinder SE detection. Later, at 80 ps, the density of the excess carriers reaches its maximum. On a longer time scale, the diode is relaxed back into the equilibrium state across the junction via recombination on the nanosecond time scale. This imaging of the charge dynamics with high spatial and temporal resolutions by SUEM paved the way for direct imaging of fast dynamics in more complex systems. Similarly, the photo-induced charge diffusion, together with the spontaneous electron–hole separation and charge trapping induced by atomic disorder, were studied in hydrogenated amorphous silicon (a-Si:H) in [108].

*5.5. Direct imaging of transient structural dynamics and 4D electron tomography*

The development of UEM helps in connecting the ultrafast electron diffraction dynamics measurement with the structure and morphology of the sample under study. Barwick et al. [81] utilized UEM to study the structural dynamics and morphological changes in single-crystal gold and graphite films. In this study, the structural change of a specific area of the sample was induced using an ultrafast laser pulse through local heating while the image frames and diffraction patterns were recorded at different instants of time. A single-crystal thin film (11 nm) of gold was first illuminated with the laser pulse (fluence of 1.7 mJ/cm$^2$). Bright-field images of a specific area of the sample were captured at a different time delay. Then, the sample was tilted to form an angle of 10° with the microscope axis. The difference between the recorded images and a reference image (at the absence of the pump pulse) was obtained. These acquired images showed the structural change of the specific area of the sample (Fig. 8(a)). By using the cross-correlation imaging method, the dynamics curve was retrieved, which showed two time constants: one at 90 ps and another at 1 ns. In addition, the electron diffraction pattern was recorded at negative and positive time delays to confirm this structural change. Moreover, in this work, the graphite structural change was studied by recording off-axis (at an angle of 21° to the microscope axis) selected-area images. Coherent resonance modulations with a resonance time of 56.3 ± 1 ps in the image (and in the diffraction pattern) were directly observed and imaged in real time. However, the decay of the envelope for this particular resonance occurs at 280 ps, which indicates that the change in the film thickness (150 nm) is on the order of ±3 nm.

Additionally, UEM allowed recording of a sequence of time frames representing a complete tilt series of 2D projections of the sample. Different time frames of the tomograms constitute a movie of the object in action, which enables the study of non-equilibrium structures and transient processes. Kwon et al. [109] demonstrated 4D electron tomography by recording different modes of motion, such as breathing and wiggling of a bracelet-like ring structure of carbon nanotubes with resonance frequencies of up to 30 MHz. Figure 8(b) shows these different types of motion of the carbon nanotube triggered by a laser pulse at different time scales (pico- to nanosecond), not limited by the recording rate of the CCD camera.

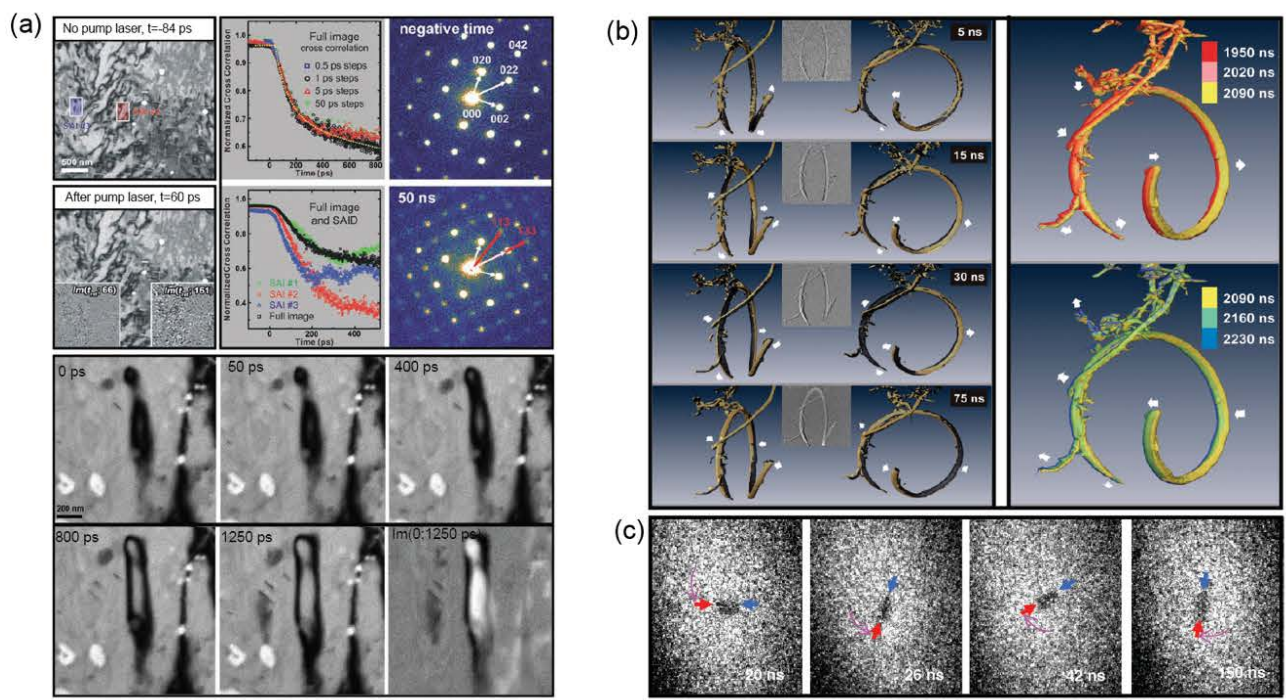

Figure 8. Imaging the ultrafast dynamics in UEM. (a) Direct imaging and electron diffraction of ultrafast structure dynamics and morphological change of single crystal gold film [81]. (b) 4D electron tomography and imaging of carbon nanotube motion induced by laser heating in real time and space [109]. (c) Imaging of the nanosecond rotational dynamics of gold nanodimer in the liquid state [84].

Recently, as reported in [84], we were able to implement a liquid cell in UEM, which allowed us to the study and image the photo-induced rotational motion of gold nanoparticles and its connection to the particle morphology in the liquid phase. We illuminated a dimer consisting of two NPs with diameters of 57 and 66 nm floating in an aqueous solution capped in the liquid-cell structure by a single femtosecond laser pulse with a fluence of 10 mJ/cm$^2$. Single-shot images were then recorded at different delay times. The dimer orientation changed by specific angles with respect to the initial state at different delays. Therefore, the relative rotation angle increased with the delay time (Fig. 8(c)). The retrieved rotation angles were 0°, 2°, 12°, 17°, 22°, and 29° at time delays of 10, 20, 26, 42, 90, and 150 ns, respectively. The random rotation angles in both the clockwise and anticlockwise directions indicate that the rotational dynamics of the dimer is ballistic and occurs on the nanosecond time scale. The retrieved rotation angle increased only to 2° between 10 to 20 ns, while it increased rapidly to 17° at 42 ns, followed by a very slow increase. The rotation angle at the delay of 150 ns (29°) is nearly 90% of the total rotation angle (33°) observed after the excitation pulse. This work represents the first study on the ultrafast dynamics in the liquid phase inside the UEM, which opens up the door for many applications to study the structural dynamics of biological molecules in their native environments.

*5.6. Biological applications*

As discussed earlier, UED and UEM have a variety of applications in chemistry, physics, and materials science. Similarly, these techniques are used to provide the real-time dynamics of biological systems such as protein vesicles, bacteria, amyloid proteins, and DNA nanostructures. In this section, we will discuss some of these applications.

Recently, it became possible to image the protein and cellar structures in their native environment by utilizing the Cryo-Electron Microscopy (cryo-EM) technique, where the sample is placed in frozen water (glassy ice) such that the sample image is not affected, as in the case of using ordinary crystalline ice. The cryo-EM technique was adopted in UEM to study the photo-induced dynamics of biomolecules in the hydrate state in [110]. In this study, the picometer-scale movements of a thin film of photoresponsive insulin amyloid fibrils embedded in vitreous ice were detected on a nanosecond time scale. A small amyloidophilic dye molecule (Congo red) was bound to the outer surface of the amyloid fibrils to increase the photoabsorption of visible photons. The nanosecond probe electron pulse (120 KeV) in the UEM was generated via the photoemission process by focusing a nanosecond UV pulse on the photocathode. Visible laser pulses of ∼ 6 mJ/cm$^2$ fluence at a 1 kHz repetition rate then illuminated the sample, leading to the expansion of the fibrils due to thermal heating. The expansion movement was probed by recording a series of electron diffraction frames of the fiber at different delay times ranging from 100 to 500 ns in 50 ps increments. This demonstration paved the way for the study of time-resolved dynamics of more complex biomolecules such as protein crystals in similar environments. Accordingly, researchers used the same technique to visualize the oscillatory dynamics of individual freestanding amyloid nanocrystals directly and noninvasively [101, 111]. On the other hand, the mechanical properties of a freestanding DNA nanostructure were visualized by UEM in [112]. Moreover, the development of PINEM allowed direct imaging of different biological materials [113], which will be discussed in detail later in the section on PINEM.

## 6. Electron source development

### 6.1. Electron pulse generation

The development of the electron source and the generation of ultrafast electron pulses played a vital role in the establishment of ultrafast electron diffraction and microscopy. The developments that are being carried out on the electron source focus on the enhancement of the temporal and spatial coherent resolution to capture faster dynamics of matter with higher resolution.

Spatial coherence, which is defined through the maximum distance between two objects for which interference is still visible on the screen, is a critical parameter of the electron source and is determined by the transverse and longitudinal coherence lengths of the source (see Section 3). On the other hand, the temporal resolution is defined by the electron pulse duration and the phase stability at the interaction stage. Both the temporal and spatial resolutions are reduced by the space-charge effect due to Coulomb repulsion of electrons in the wave packet [114], as discussed earlier in Section 2. The electron beam intensity requirements for the source are more difficult to define, since they depend on the sample under study, namely on the phase (i.e., solid, gas), the thickness, which should be in the range of 10–100 nm, and the background. Nevertheless, at least $\sim 10^4$–$10^5$ electrons per pulse are needed to clearly resolve the diffraction peak and/or obtain a clear image in UED and UEM, respectively [18]. From the theoretical modelling and experimental study, it is obvious that it is impossible to use the conventional electron gun parameters to attain sub-ps electron pulses at the sample position with reasonable electron number densities [114-116]. In the pioneering work by Siwick and Miller on implementing a compact electron gun design, 600 fs pulse durations with up to 10,000 electrons per pulse were achieved [22]. This paved the way for the development of a source of high-brightness electron beams with pulse durations on the order of hundred femtoseconds, based on Radio Frequency (RF) compression [96, 102, 117], which will be discussed later.

### 6.2. Electron pulse compression techniques

RF electron pulse compression (schematic layout shown in Fig. 9), which is based on inverting the linear energy chirp and temporally focusing the electron pulses, has been theoretically studied and modelled elsewhere [116]. It was demonstrated by Van Oudheusden et al. [96] for controlling the space-charge effect and generating ultrashort electron pulses. In this demonstration, the design of the RF cavity has a cylindrical $TM_{010}$ RF pill box, which is capable of recompressing the electron bursts by providing an axial electric field. This field is uniform along the cavity axis and varies in time according to the following equation:

$$E_z(z,t) = E_0(z)\cos(2\pi f\, t + \phi),$$

where $f \approx 3$ GHz is the resonant frequency of the cavity (this is the standard value of S-band technology used by the RF accelerator community) and $\phi$ is the phase, which is chosen to be zero when the center of the electron pulse crosses the center of the cavity in order to ensure that there is no overall gain or loss of energy. The RF compression process (Fig. 9) can be explained as follows. The electron pulse is generated from the photocathode. The electron beam is then focused by a magnetic lens, and the longitudinal momentum–position distribution becomes linear prior to entering the RF pill lens. Inside the RF pill box, the applied field affects the electron pulse, and thus the faster electrons are pushed backward while the slower electrons are pushed forward. Therefore, the output pulse is positively linearly chirped, and the longitudinal momentum–position distribution is reversed. The sample position is accurately defined to be at the longitudinal focal point to ensure that the slower and faster electrons meet at that position to obtain the maximum temporal confinement of the electron pulse. Magnetic lenses are used to form the electron beam in the transverse direction, which in turn leads to an increase in the spatial resolution due to the longer path length to the sample and the smaller "local" transverse momentum spread [18, 118, 119]. This scheme has been used to produce 70 fs electron pulses with an areal density of $2.5 \times 10^8$ electrons cm$^{-2}$ [96]. These ultrashort high-brightness electron pulses made it possible to visualize the fast atomic motion in real time utilizing UED [103]. Many studies based on the same idea have been reported. These studies aim to compress the electron bunches to several femtoseconds by utilizing different approaches such as microwave [120-123], terahertz [124], and DC electric fields [125-127].

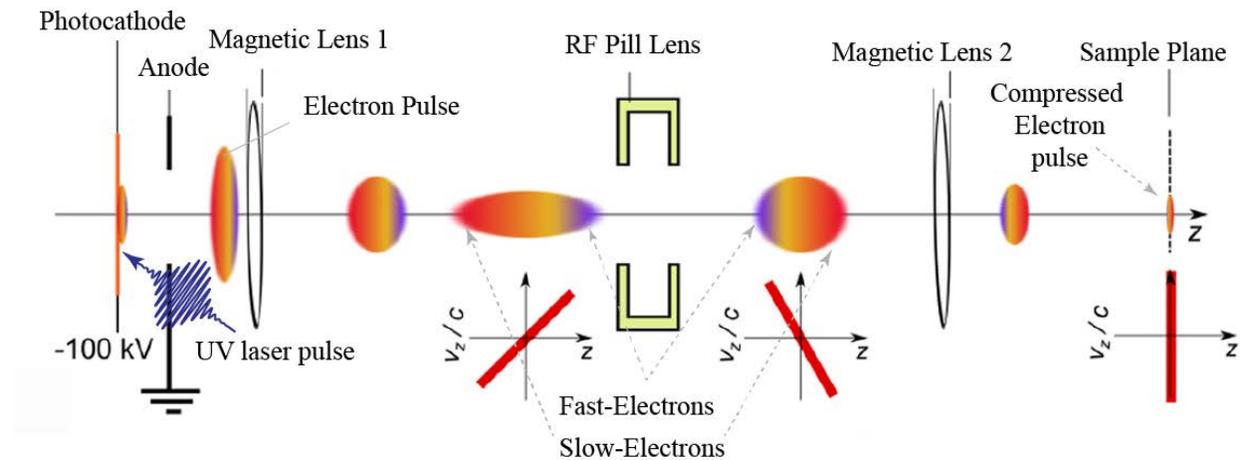

Figure 9. Layout of the RF compression scheme adopted from [18].

The main challenge in these approaches is the synchronization and phase jittering between the compressed electron pulse and the trigger (laser) pulse in the time-resolved measurements, which limits the temporal response of the UED apparatus. Therefore, the RF field phase has to be locked to the pump laser pulse to avoid pulse-to-pulse time jittering. Previously, in time-resolved experiments, phase-locking has been implemented using external voltage-controlled oscillators [97, 120] or repetition rate multiplication techniques [128]. In both approaches, the time jittering stability depends on the pulse energy stability of the laser and is limited by the amplitude–phase conversion. Due to the relatively long acquisition time of the time-resolved measurements (several hours) and the desired precise synchronization between the pump and probe pulses, this synchronization issue limits the temporal resolution to hundreds of femtoseconds, preventing the access to faster dynamics beyond that limit. Recently, Otto et al. [129] demonstrated an important technical improvement and enhancement in microwave active phase synchronization. This enhancement was based on the elimination of several sources of phase instability inherent in previous approaches, such as (i) amplitude phase errors in the generation of the microwave signal, (ii) phase instability in power amplification, and (iii) phase drift in the cavity response due to thermally-induced frequency detuning. The reported stability is on the order of ~ 10 fs, and the long-term arrival time stability of the electron pulse is < 50 fs. For more technical details, please refer to [129]. We expect that this advance will play a vital role in enhancing the temporal resolution for resolving faster atomic motion.

Various approaches have been reported for the electron source development and generation of ultrafast electron pulses. Maxson et al. [130] demonstrated the generation and characterization of < 10 fs electron pulses with $10^5$ electrons/pulse. This was achieved by using a special configuration of an RF compressor where a ∼ 100 fs laser pulse is tightly focused onto the cathode of a 1.6 cell RF photo injector and employing a velocity-bunching cavity to compress the electron bunch. However, this has not been utilized yet in time-resolved diffraction or imaging measurements. Similarly, there are many research activities focused on improving the electron source and electron optics inside the microscope to achieve better temporal and spatial resolutions. Feist et al. [131] developed a UEM electron source based on laser-triggered electron emission from a nanoscale photocathode, which enables 200 fs temporal resolution and an ability to focus the photoelectron beam to the sub-nm scale. This has been done by employing a laser-driven Schottky emitter, which is placed inside an electrostatic suppressor–extractor electrode assembly. It is illuminated by a 400 nm laser beam focused down to 20 μm (beam diameter) to generate the electron pulse that can be accelerated to 200 keV and utilized for imaging the sample [131]. Moreover, some other schemes based on the implementation of a nanotip as an electron source to achieve femtosecond and nanometer spatiotemporal resolution have been proposed [132-138]. Alternatively, C.Y. Ruan and Co. are working on developing a UEM with high-brightness femtosecond electron beams based on tuning the electron optics to maintain both the temporal and spatial resolutions and minimize the space-charge effect. This work uses a new type of energy filter for condensing the energy spread of the electron pulses to the emittance-limited width without sacrificing the electron count. This will have a significant impact for single-shot ultrafast imaging and high-resolution electron spectroscopy measurements in the next generation of ultrafast electron microscopes [139]. Recently, Hassan and co-workers [26] demonstrated the highest temporal resolution that has been achieved in electron microscopy, which breaks the limits by attaining a 30 fs temporal resolution using the optical gating approach based on PINEM. This provides the required resolution for imaging the ultrafast electron dynamics for the first time in UEM. This promising approach has a great potential to generate attosecond or femtosecond electron pulses and bring the temporal resolution to the attosecond time scale, which will be discussed in detail in the next section.

Although the enhancement of temporal resolution is a crucial aspect of the electron source development, the enhancement of spatial resolution has also attracted the attention of many groups at the national facilities, as we will illustrate next.

**Mega-electron-volt ultrafast electron diffraction (MeV-UED) at SLAC National Accelerator Laboratory**

The second generation of UEDs with mega-electron-volt (MeV) electron sources have been developed recently at SLAC [43]. This development has several advantages. First, the MeV electron pulse energy reduces the space-charge effect and consequently maintains the temporal confinement and brightness of the electron beam during propagation to the sample stage. Second, utilizing the MeV electron pulse overcomes the loss of temporal resolution due to the velocity mismatch issue of the pump laser pulse. Third, it enables efficient acquisition of data with high signal-to-noise ratio within a reasonable exposure time. Finally, it can be used to study both gas- and solid-phase samples. For these reasons, this powerful tool is promising for the imaging of ultrafast dynamics in chemical and biological complex systems. As some of the first applications, MeV-UED has been used to study and image the rotational wave packet dynamics of nitrogen [44] and isolated iodine molecules [99] with 230 fs temporal resolution and sub-Angstrom (0.76 Å) spatial resolution.

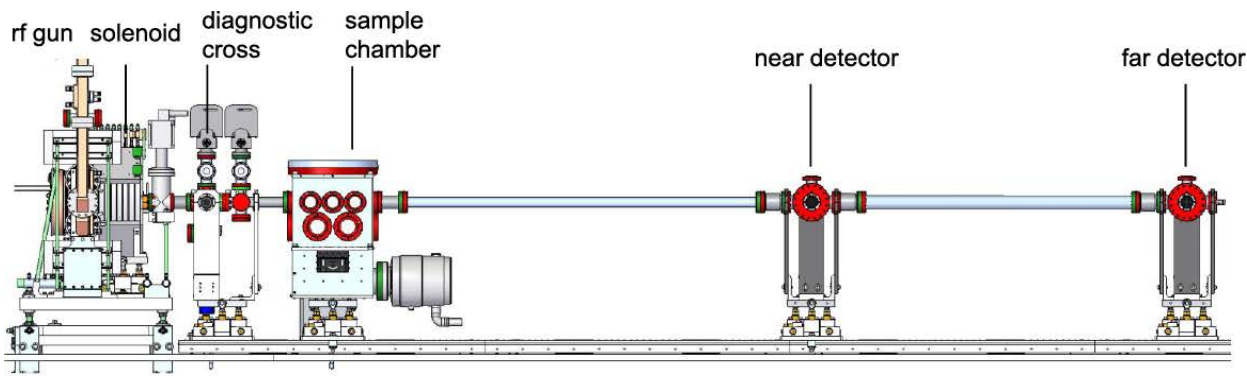

Figure 10. Schematic of mega-electron-volt ultrafast electron diffraction (MeV-UED) at SLAC [43].

A schematic of the MeV-UED beamline at SLAC is shown in Fig. 10 [43]. The system includes a Linac Coherent Light Source (LCLS) type photocathode RF gun, a sample chamber, a high-efficiency electron detector, an ultra-stable RF power source, a Ti:Sapphire laser, and a laser–RF timing system. Importantly, a closed timing loop was implemented to achieve the minimum time jitter between the pump laser and probe electron pulse. The basic setup for the time-resolved diffraction experiment is similar to the conventional UED experimental setup; more details can be found in [43]. Currently, new work is being carried out at Lawrence Berkeley National Laboratory (Berkeley Lab) to establish the High Repetition Rate Electron Scattering (HiRES) apparatus by Daniele Filippetto and co-workers. The main goal is to generate high-intensity femtosecond electron pulses (106 electrons/pulse) with MHz repetition rates. The implementation of this electron source in UED holds great potential for different types of applications [140].

## 7. Photon–electron interaction: optical gating of electron pulses

### 7.1. Photon-Induced Near-Field Electron Microscopy

PINEM, a key UEM technique, is based on the photon–electron interaction [82]. The basic principle of PINEM can be explained as follows. In free space, an electron cannot absorb a quantum of electromagnetic energy because of the lack of energy–momentum conservation. However, in the presence of a nanostructure, inelastic coupling between free electrons and photons takes place [141, 142] due to the deceleration of the scattered photons and satisfaction of the energy–momentum conservation condition. The coupling leads to gain/loss of photon quanta by electrons in the electron packet, which can be resolved in the electron energy spectrum [82, 143-145]. This spectrum consists of discrete peaks, spectrally separated by multiples of the photon energy ($n\hbar\omega$), on the higher and lower energy sides of the zero loss peak (ZLP) [82] (see Fig. 11). The development of PINEM opens up the door for various applications such as the imaging of biological structures [113], visualization of plasmonic fields [82, 145] and their spatial interference [146], visualization of the spatiotemporal dielectric response of nanostructures [147], imaging of low-atomic-number nanoscale materials [148], and characterization of ultrashort electron packets [149, 150].

In this section, we will discuss some of these applications. In biology, PINEM has been used to image different biological structures such as protein vesicles and whole cells of Escherichia coli [113]. In this work, the electrons gain energy due to the photon–electron coupling, with the laser pulses filtered out and these biological structures illuminated. In the dark field images (Fig. 12(a)), these structures are lighted up and enhanced. The contrast enhancement is controlled via the laser polarization, time resolution, and tomographic tilting. This work opens the way for a variety of biological PINEM applications such as imaging and identifying the cancer cells, as recently reported in [151]. Moreover,

PINEM has been used to image confined static plasmonic fields in real space, as reported by Piazza et al. [146]. The authors reported imaging of a photo-induced surface plasmonic standing wave on a metallic nanowire (Ag nanowire) (Fig. 12(b)). In addition, the control of spatial interference of the excitation plasmonic field was demonstrated. Accordingly, the cross-correlation images of the excited surface plasmon were obtained by controlling the relative delay between the driver laser pulse and electron pulse.

On the other hand, the indirect PINEM imaging (spectral mapping) enables visualization of the excited surface plasmons on a sub-particle scale [147]. This is done by focusing the electron beam onto a single nanoparticle and recording the PINEM spectra at each spot on the particle surface and then scanning over the particle. This is repeated at different relative delays between the electron and optical pulse to obtain a series of spectral mapping images as shown in Fig. 12(c). PINEM has also been used to implement coherent quantum control of the free electron population state, as demonstrated by Feist et al. [152]. This was done by controlling the photo-induced Rabi oscillations in the populations of electron momentum states via changing the intensity of the optical driving field (Fig. 12(d)). This work was conducted on a conical gold tip. The interaction of the electron and optical pulses leads to the generation of PINEM peaks. The number of PINEM peaks reflects the interaction process between the optical pulse and gold nanotip. Altering the driving field enhances the population of the PINEM peaks linearly. The results showed that modulations are directly related to the multi-level Rabi oscillations and thereby quantum coherent manipulation of the respective level amplitudes (Fig. 12(d)) [152]. In addition, the generation of attosecond pulse trains in UEM has been anticipated by exploiting the interference between two induced near-fields at a certain propagation distance from the coupling interaction stage. However, the realization of these pulses and their isolation remain technically very challenging in time-resolved electron microscopy experiments [152, 153].

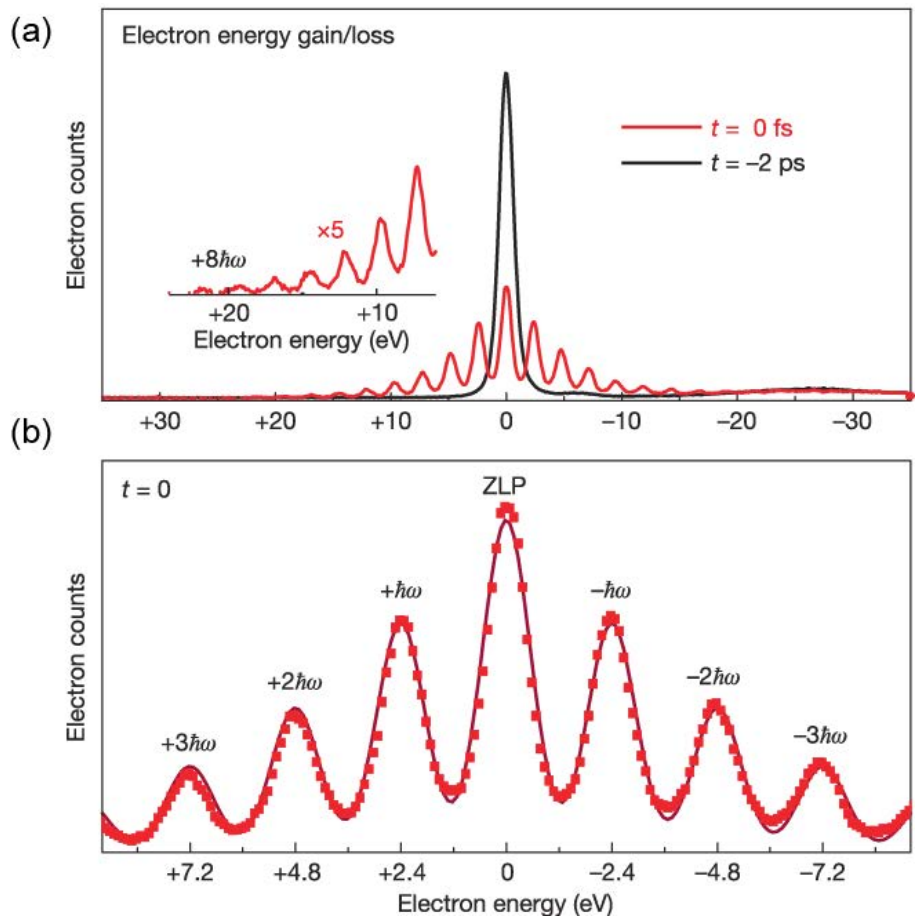

Figure 11. Photon-Induced Near-field Electron Microscopy (PINEM). (a) The electron energy spectrum in the absence of an optical pulse (ZLP) is plotted in black, while the spectrum of the electron-photon coupling at the temporal overlap (t= 0 fs) between the optical and electron pulses is shown in red. The magnification of this coupling electron energy spectrum showing the separate peaks of PINEM on both sides of ZLP is shown in (b) [82].

Another important application of PINEM is the characterization of the temporal profile of ultrafast electron pulses, which has been demonstrated by Kirchner et al. [149]. They invoked the basic principle of the attosecond streaking camera [154, 155] to demonstrate the optical-field-driven streak camera in electron microscopy. This technique is considered to be one of the most powerful approaches to temporally characterize both the electron and laser pulses (in the case of extremely short electron pulses) involved in the PINEM coupling with sub-femtosecond resolution. In the attosecond streaking, the XUV pulse releases an electron wave packet by photoemission, and momentum exchange occurs in the presence of the optical field. Both the optical field and temporal profile of the attosecond XUV pulse can be retrieved by tracing the momentum change of the electron wave packet [155]. Likewise, the energy exchange between the free electron pulse and optical pulse in PINEM is the basis of light-field-induced free-electron streaking reported in that work. In Fig. 12(e), the calculated streaking spectrogram for 50 fs laser pulses and 300 fs electron pulses, with a chirp of 1.6 eV, are shown in case (1), while the spectrograms of a phase-locked 5 fs laser pulse and electron pulses with durations of 3 fs and 1 fs are shown in cases (2) and (3), respectively. Eventually, this demonstrates that the light-field-induced free-electron streaking could be utilized for the temporal characterization of attosecond electron pulses.

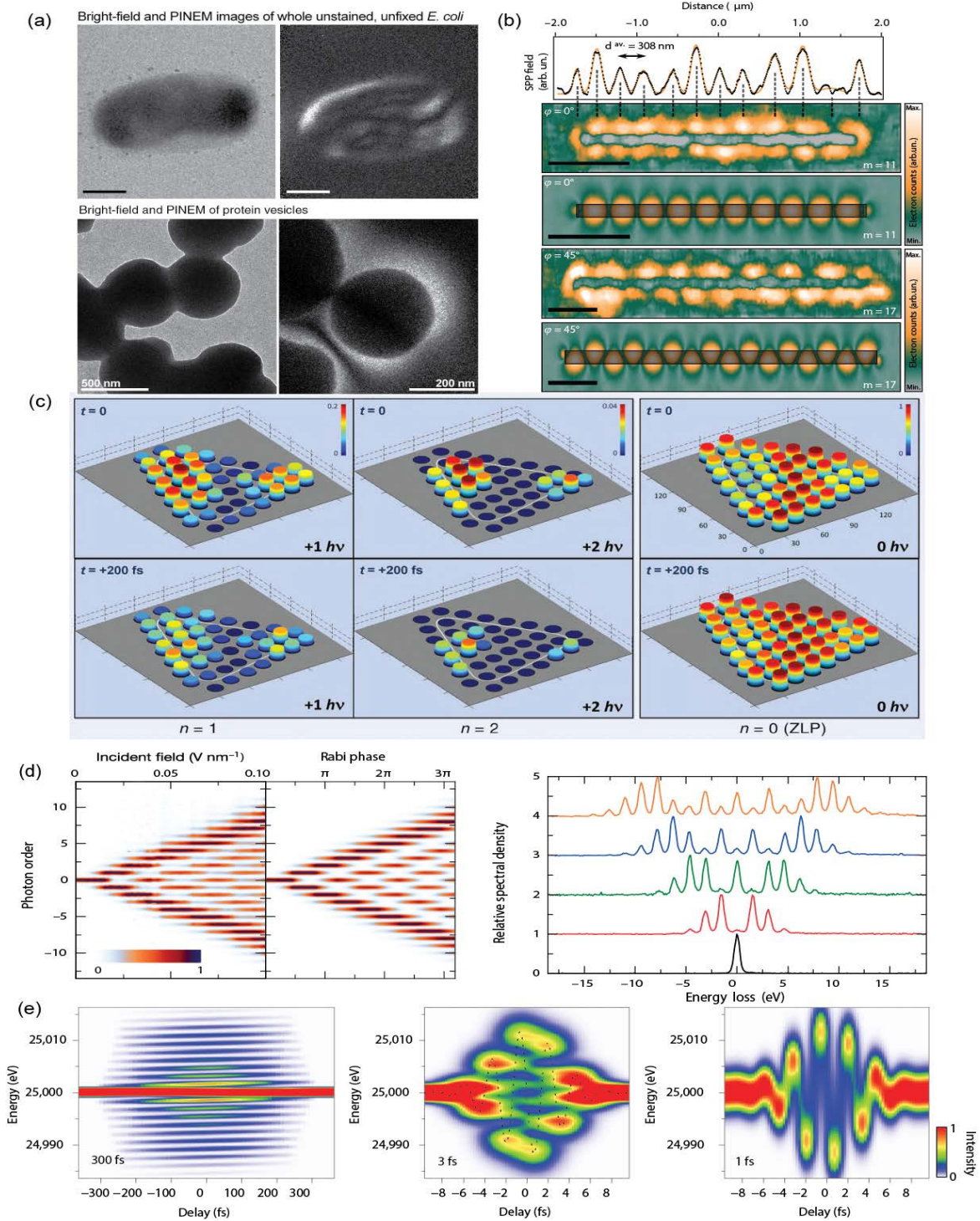

Figure 12. Photon-Induced Near-field Electron Microscopy (PINEM) applications. (a) PINEM imaging of biological structure (protein vesicles and whole cells of Escherichia coli) [113]. Imaging of the excited plasmonic field on silver nanowire [146], and the surface of silver triangular nanostructure [147] is shown in (b), and (c) respectively. (d) The coherent quantum control of the free electron population state probed by PINEM [152]. (e) Laser field streaking of free electron pulse [149].

Accordingly, Kozák et al. [156] demonstrated laser-driven gating and streaking of the free electron pulse based on the interaction between electrons and the photon-induced near field. The optical interference between two laser pulses coming from the same laser source ($\lambda$ = 1.93 µm, duration of 630 fs) allowed for the trapping and temporal gating of the electron gain/loss energy due to the coupling with the laser pulse in a sub-optical-cycle (6.5 fs) temporal window. The control of the relative delay between the two driver laser pulses offered control on the temporal resolution on the order of 1.2 ± 0.3 fs [156]. In these previous applications, the PINEM technique was exploited to characterize the temporal profile of the electron pulse and to image the induced near field of the system in the static state. Next, we will discuss the first application that utilized PINEM to study the ultrafast dynamics of matter.

*7.2. Time-resolved Photon-Induced Near-Field Electron Microscopy (TR-PINEM)*

The basic principle in time-resolved PINEM measurements, reported in the study by Hassan et al. [106], can be explained as follows: the PINEM signal is produced due to the coupling between the electron pulse and the first laser pulse (both pulses are kept in spatiotemporal overlap during the experiment). Then, the electrons that gain and/or lose photon quanta (represented by the PINEM peaks) are used to "probe" the ultrafast dynamics of matter triggered by the second laser "pump" pulse.

In this work, we exploited time-resolved PINEM to study the ultrafast phase transition dynamics of $VO_2$ nanoparticles from the initial (monoclinic) insulator phase to the (tetragonal) metal phase [106]. The photo-induced dielectric response of $VO_2$—which is strongly related to the lattice symmetry [157]—manifests itself in the change of the PINEM peaks intensities. Therefore, the ultrafast phase transition dynamics were retrieved by tracing the changes of these PINEM peak intensities as functions of the pump laser pulse delay.

The experimental setup is shown in Fig. 13, and the principle is illustrated in the inset. First, an infrared laser pulse (250 fs) with a central wavelength of $\lambda \sim 1039$ nm is frequency doubled to generate a visible laser pulse (~ 519 nm). The visible laser beam is divided into two beams by a dichroic beamsplitter. The first visible laser beam is utilized to generate DUV laser pulses ($\lambda \sim 259$ nm) through the second harmonic generation process. These pulses are directed to the photoemissive cathode (inside the microscope) to generate ultrafast electron pulses, which are accelerated in the column (~ 200 keV). The second visible laser beam is divided into two pulses by a second beamsplitter. The first visible laser pulse ($P_1$) couples with the electron pulses inside the microscope in the presence of the specimen ($VO_2$ nanoparticles) and generates the PINEM peaks, while the two pulses are maintained at the spatiotemporal overlap. The second visible laser pulse ($P_2$) acts as a pump pulse, and induces phase transition in the $VO_2$ nanoparticles. The delay of the pump pulse with respect to the electron pulse ($\tau_2$) is controlled by another precise linear delay stage.

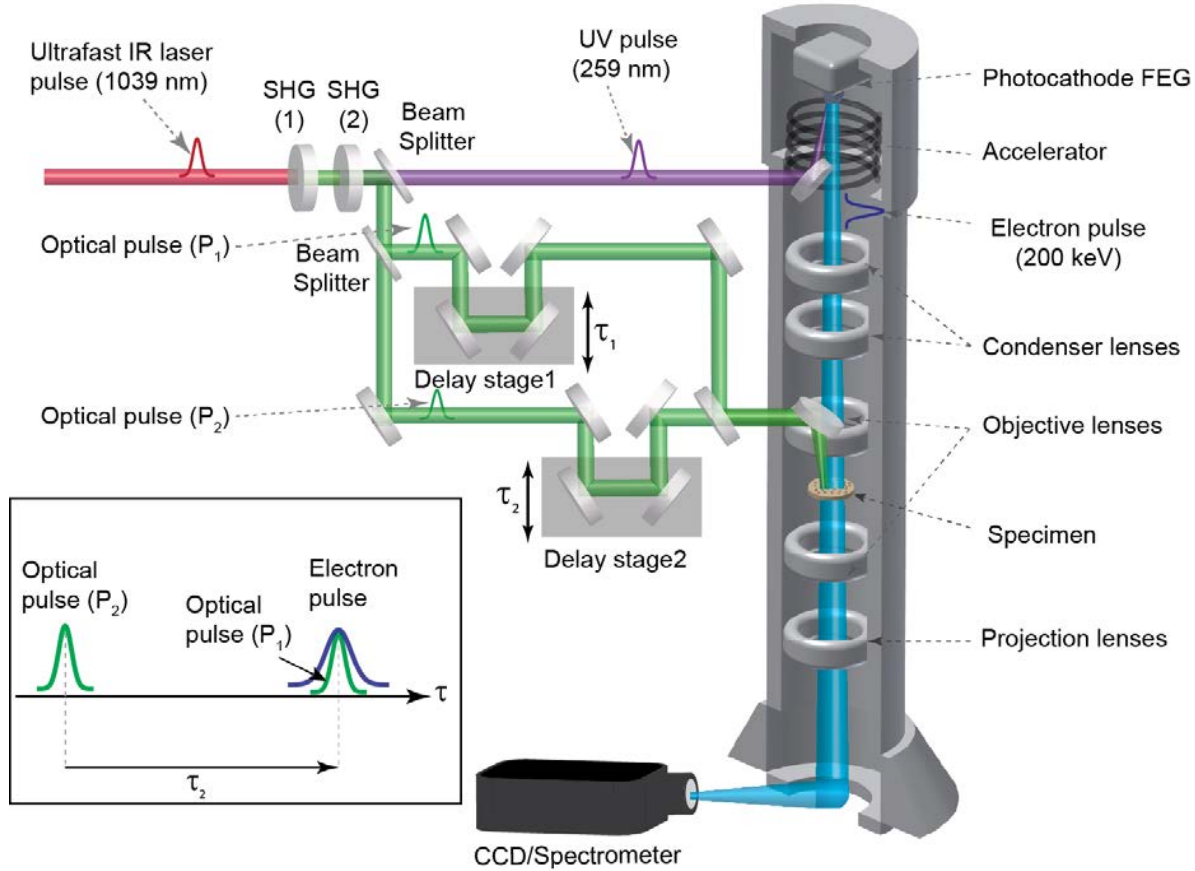

Figure 13. Time-resolved Photon-Induced Near-field Electron Microscopy (PINEM) experiment setup and principle.

The change in the intensity of the PINEM spectrum due to the photo-induced insulator–metal phase transition was traced (Fig. 14). Here, each data point represents the integration of the PINEM spectrum at a certain delay time ($\tau_2$). This change reflects the dielectric response of $VO_2$ during the phase transition process. The PINEM spectrum intensity remains unchanged until the arrival of the pump pulse, when the phase transition of the $VO_2$ nanoparticles starts taking place. The dielectric properties of the nanoparticles are modified due to the change in the lattice structure during the phase transition [157]. A biexponential fitting of the PINEM intensity dynamic curve (red line, Fig. 14) reveals two time constants (10 ps and ~ 170 ps); this is attributed to the vanadium atom motion within the unit cell and long-range shear rearrangement that is essential for the rutile phase transformation process [158]. To confirm and validate this measurement, a conventional time-resolved electron diffraction measurement was conducted on the same specimen under the same conditions. The results are in acceptable agreement, which demonstrates the validity of the time-resolved PINEM measurements [106]. This work opens the way for exploiting the PINEM technique to study the dynamics of matter in real time and space.

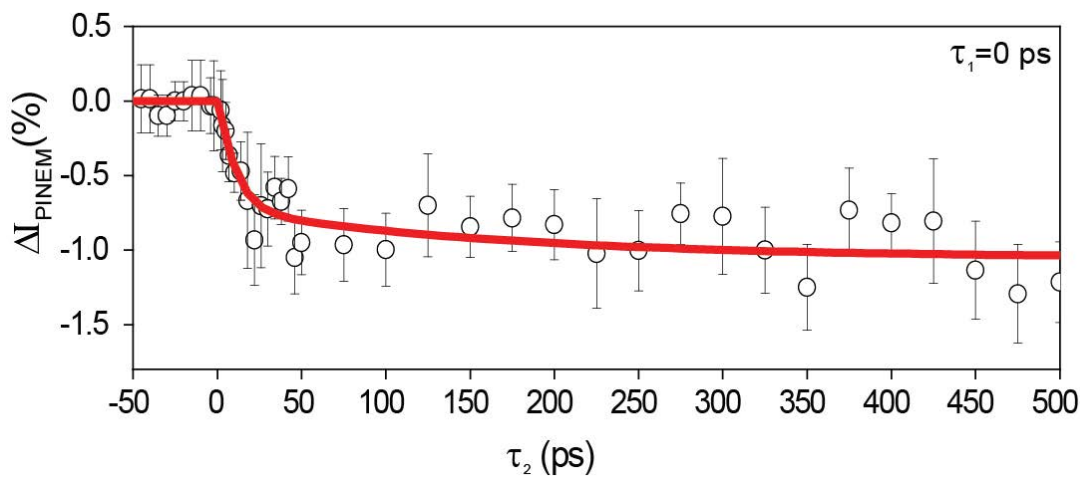

Figure 14. Study of the phase transition in $VO_2$ nanoparticles by time-resolved PINEM. The relative change in the PINEM spectrum intensity as a function of the second optical pulse (pump) delay ($\tau_2$) is plotted using the black open circle points, where each data point represents the integration of the PINEM spectrum at time ($\tau_2$). The red line is the biexponential fitting of the measured dynamics curve.

*7.3. Optical gating of ultrafast electron pulses*

As explained earlier, some of the electrons in the wave packet gain or lose multiple photon quanta only in the presence of the optical laser pulse, due to the photon–electron coupling in PINEM. This means that the optical pulse acts as a "temporal gate" for these electrons [106, 159]. These gated electrons have a temporal profile that emulates the gating window (i.e., the optical pulse duration). It can be filtered out to obtain ultrashort electron pulses, thus providing significant enhancement of the temporal resolution in electron microscopy for exploring ultrafast dynamics of matter triggered by other ultrashort optical laser pulses in different UEM modes (i.e., diffraction, electron spectroscopy, and/or direct imaging) [82, 106, 152, 160, 161]. The current advancements in attosecond physics and the generation of optical pulses lasting a few hundreds of attoseconds [27], which can be utilized as gating pulses, might lead to the generation of few-femtosecond and/or attosecond electron pulses, as we will explain in the next section.

*7.4. Shortest isolated electron pulse in UEM*

The temporal resolution in UEM mainly depends on the duration of the electron pulse, which is finite due to the initial energy dispersion and the space-charge effect. In UEM, the typical temporal resolution spans the range from several hundreds of femtoseconds to a few tens of picoseconds, which is insufficient to resolve the faster transient dynamics of matter lasting from few tens to few hundreds of femtoseconds. Hence, the generation of short (few tens of femtoseconds) electron pulses is highly desirable for enhancing the temporal resolution of UEM.

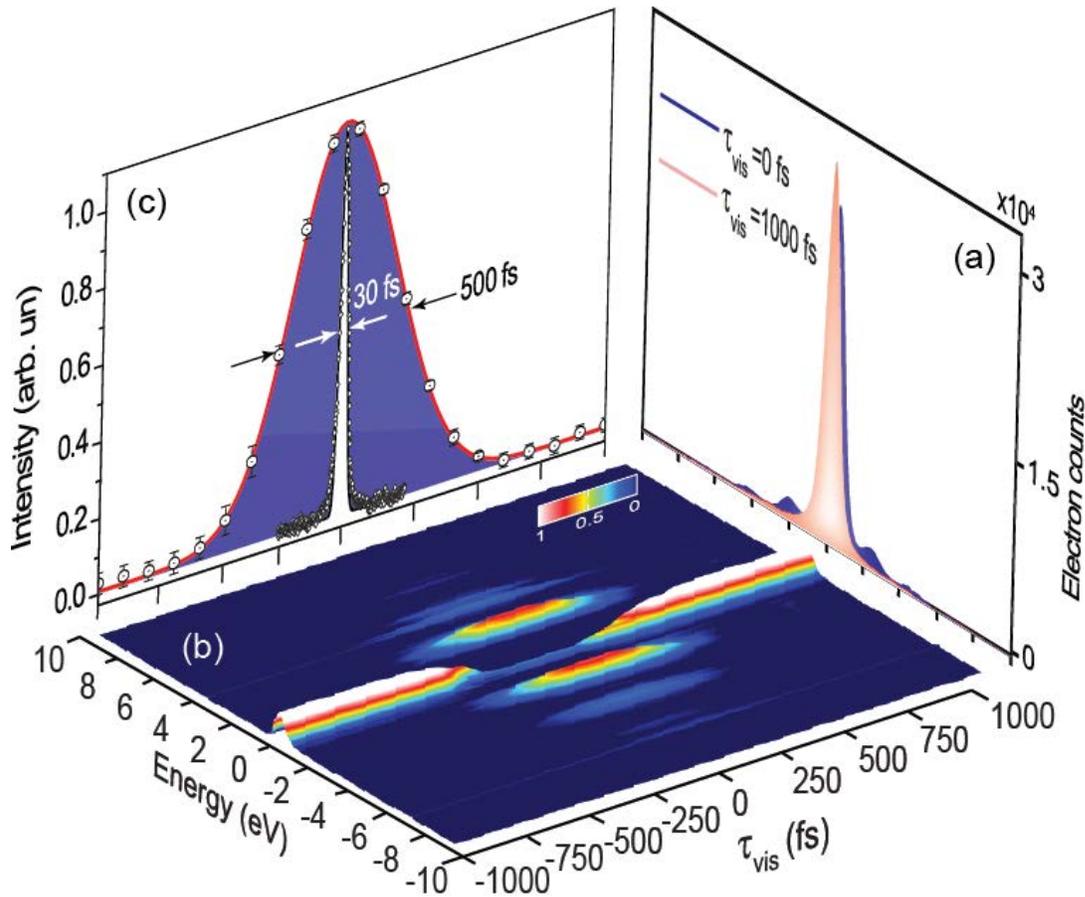

Figure 15. Characterization and optical gating of ultrafast electron pulse in UEM. (a) The electron energy spectrum of the "original" ultrafast electron pulse (ZLP spectrum) is shown as a reddish white curve, and the coupling between the electron and the visible "gating" (ħω=2.25eV) laser pulse is shown as a blue curve. (b) Measured spectrogram of the electron energy spectra as a function of the "gating" visible laser pulse delay $\tau_{vis}$. (c) The cross-correlation temporal profile is shown as open black circles, and its fitting is indicated by the blue shaded region and red curve; the cross-correlation profile has an FWHM of 500 fs. The white shaded curve and white dotted line represent the measured temporal gating window (the pulse duration of the "gating" visible pulse).

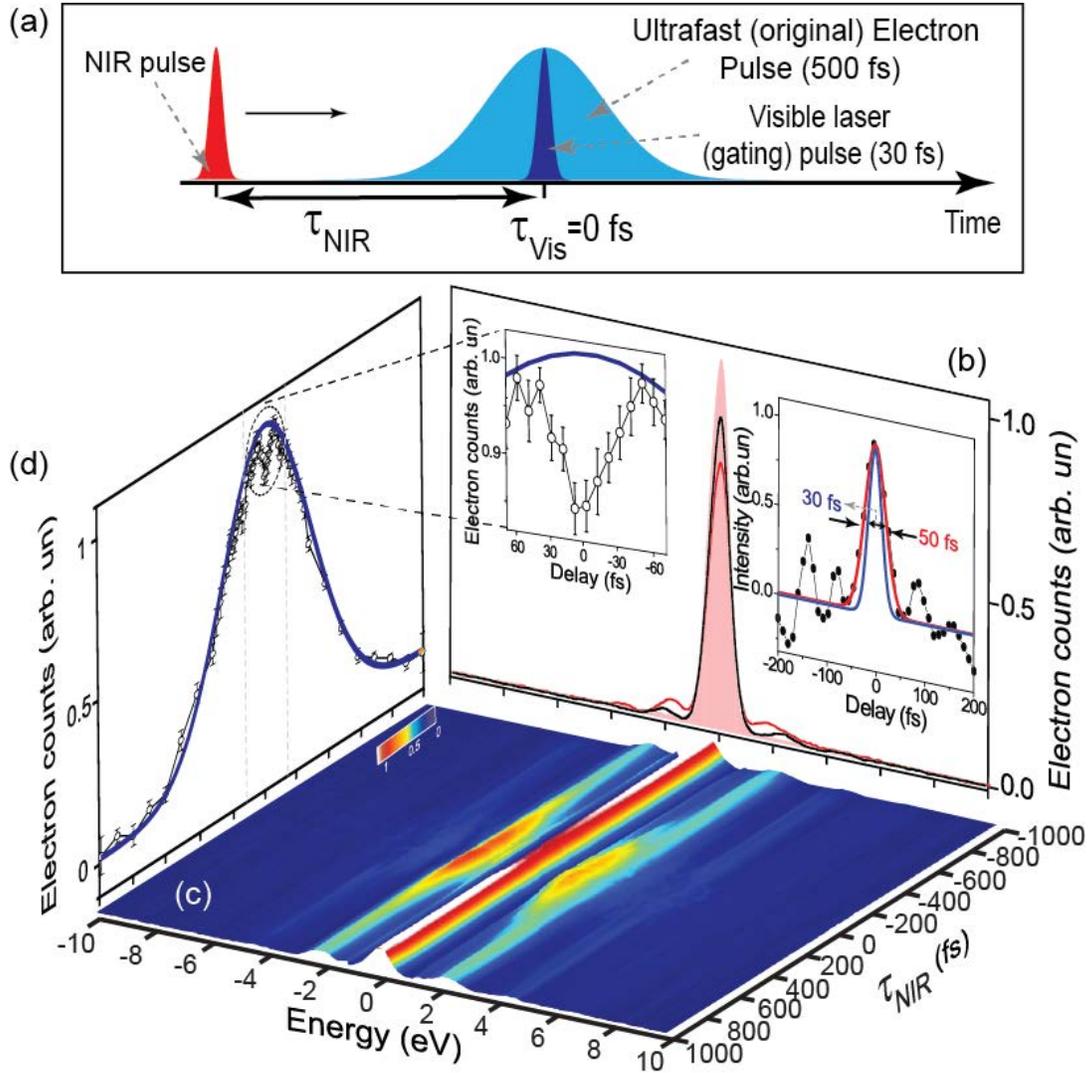

Figure 16. Temporal characterization of the gated isolated 30 fs electron pulse. (a) Illustration of the gated electron pulse temporal characterization principle based on the cross-correlation measurement. (b) The electron energy spectra of the coupling between "original" electron pulse and "gating" visible pulse is represented by the black line, while both the visible and NIR ($\hbar\omega=1.675$ eV) laser pulses are represented by the red line. The spectrum of the "original" electron pulse (ZLP) is indicated by the white reddish curve. (c) Cross-correlation electron energy spectrogram of the electron-photon coupling between the NIR laser pulse and both the "original" and "gated" electron pulses. The ZLP is suppressed to obtain a clear illustration of the gating effect. (d) The cross-correlation temporal profile retrieved from the measured spectrogram is shown in (c) by the connected open circles. This curve and its expanded view (inset in (b) left) clearly show the dip in the electron counts due to the "gated" electron pulse. The right inset in (b) shows the cross-correlation temporal profile of the "gated" electron and NIR pulses, obtained by subtraction of the temporal profiles in (d). It is plotted in black dots along with its fitting (red line), and it has a FWHM in the order of 50 fs. A fit of the measured temporal profile of the "gating" window (visible pulse (30 fs)) is shown in blue.

Recently, we demonstrated the enhancement of the temporal resolution in UEM by more than an order of magnitude (16 times) by generating intense isolated ~ 30 fs electron pulses accelerated at 200 keV. These short pulses were generated by temporal optical gating. This advancement permits the resolution and imaging of fast atomic motion and electron dynamics occurring on the scale of several tens of femtoseconds.

In this work, an ultrafast electron pulse (several hundred femtoseconds) generated by photoemission inside the microscope was temporally gated utilizing a 30 fs visible laser pulse (Fig. 15). First, the temporal profile of the "original" electron pulse was characterized by cross-correlation between the "gating" visible pulse and the electron pulse. Since the "gating" pulse duration is much shorter, the cross-correlation directly reflects the temporal profile of the electron pulse. This is a common technique in electron pulse metrology [152]. The retrieved pulse duration of these "original" electron pulses is on the order of 500 fs. Second, for generating the isolated ultrashort "gated" electron pulse with maximum counts, the "gating" pulse is kept at the optimum temporal overlap ($\tau_{Vis} = 0$ fs) with the 500 fs electron pulse. Finally, the gated electron pulse, with the same temporal profile as the gating pulse, is characterized by using another laser pulse and measuring its cross-correlation temporal profile with the gated electrons. The principle of this measurement is illustrated in Fig. 16(a). The cross-correlation spectrogram (Fig. 16(c)) was recorded first. Then, the temporal profile of this cross-correlation was retrieved from the spectrogram, which carries the signature of the coupling between the original electron pulse and NIR laser pulse and that between the gated electron pulse and NIR laser pulse. The cross-correlation temporal profile of the latter is shown in the inset of Fig. 16(b), with a full width at half maximum (FWHM) on the order of 50 fs. Since the NIR pulse duration is 33 fs, the gated electron pulse duration is on the order of 30 fs, similar to the temporal profile of the gating pulse.

The generated "gated" electron pulse has sufficient electron counts (~ 8% of the total electron counts, or < 1 electron/pulse) for probing the ultrafast electron dynamics of matter. The achieved temporal resolution (30 fs) allows imaging of the electron dynamics lasting a few tens of femtoseconds, which was beyond reach before, such as that associated with the electron–electron scattering and electron–phonon coupling in semiconductors [162] and the dynamics of surface plasmons [163].

## 8. Attomicroscopy

As explained in the previous section, the optical gating approach is a promising way for generating isolated electron pulses with durations of only a few tens of femtoseconds. This approach has striking advantages over the conventional compression techniques [96, 97, 120, 124], which can be summarized as follows. First, the generated electron pulse duration is limited only by the gating laser pulse duration, which could be on the attosecond time scale [27]. It should be noted that, in this case, the nanostructure system with photon–electron coupling should have a broadband response over the spectral bandwidth of the gated pulse. Second, superior phase and synchronization stability (< 1 fs) can be achieved by active phase-locking between the two optical (pump and gating) pulses. This cannot be achieved with other electron pulse compression techniques. Therefore, this approach is considered the best candidate for achieving the attosecond resolution in electron microscopy by generating isolated attosecond electron pulses utilizing the optical attosecond pulse that we demonstrated recently [27]. In this section, we will first explain the generation of the Optical Attosecond Pulse (OAP) utilizing the light field synthesis technique. Then, we will present the theoretical calculation of the coupling between this OAP and ultrashort electron pulse for potentially generating isolated attosecond electron pulses via optical gating [26], which is the central focus of our research group activities at the University of Arizona. This will open the door for establishing the new field of "*Attomicroscopy*" and will allow femtosecond and attosecond stroboscopic imaging applications in different fields.

### 8.1. Optical attosecond pulse

The generation of an attosecond pulse spanning more than two octaves in the visible frequencies and flanking range became possible by the light field synthesis, which was demonstrated earlier [27, 164, 165]. In this process, a supercontinuum (Fig. 17) is generated by focusing multi-cycle laser pulses of 1 mJ, with a central wavelength of ~ 790 nm and a $\tau_{FWHM}$ pulse duration of ~ 22 fs, in a gas-filled Hollow Core Fiber (HCF). The Ne gas pressure inside the HCF chamber is adjusted to ~ 2.2 bar in order to maximize the bandwidth of the spectrum. This spectrum enters the light field synthesizer (shown in Fig. 18), which has been designed to synthesize the light field with attosecond resolution. The supercontinuum is divided into four nearly equal spectral bandwidths (constituent channels) utilizing Dichroic Beamsplitters (DBSs). The division of the ultrabroadband spectrum inside the synthesizer is the key point in compressing this two-octave light to the extreme limit and in generating the optical attosecond pulse. These four constituent spectral channels are: the near infrared (NIR) channel ($Ch_{NIR}$, 700–1300 nm), visible channel ($Ch_{VIS}$, 500–700 nm), visible–UV channel ($Ch_{VIS-UV}$, 350–500 nm), and deep ultraviolet channel ($Ch_{DUV}$, 270–350 nm), as shown in Fig. 17.

Inside the synthesizer apparatus, a set of six dispersive (chirped) mirrors is placed in the beam path of the corresponding channel ($CM_{NIR}$ in the NIR channel, $CM_{Vis}$ in the visible channel, $CM_{Vis-UV}$ in the visible–UV channel, and $CM_{DUV}$ in the DUV channel) to compensate for the positive dispersion of the pulse in each channel. The individual channel pulses of the synthesizer are temporally compressed close to their Fourier limit and characterized with a TG-FROG apparatus. The retrieved $\tau_{FWHM}$ pulse durations are: $\tau_{Ch(NIR)}$ = 8 fs, $\tau_{Ch(Vis)}$ = 6.4 fs, $\tau_{Ch(Vis-UV)}$ = 6.3 fs, and $\tau_{Ch(DUV)}$ = 6.5 fs, as shown in Fig. 19.

At the exit of the apparatus, the constituent channel pulses are spatiotemporally superposed with beamsplitters of the same type. Notably, there are many elements that have been implemented inside the synthesizer to control the synthesized waveforms. These elements are: a pair of movable, thin, fused silica wedges (apex: 2°48', 30 × 20 mm$^2$) introduced in the beam path of the individual channels (at the Brewster angle of each band, to minimize losses) to fine-tune the dispersion as well as the CEP phase of the constituent pulses; adjustable irises, which have been introduced in the beam path of each channel for controlling the relative intensities among the constituent channels; a translation unit, which carries a pair of mirrors in the path of each constituent channel to adjust the relative phases. This unit consists of a manually adjustable translation stage (precision ~ 10 μm), used for the coarse adjustment of the optical paths in each channel, and a piezoelectric translation stage, which is used for finer adjustments of the time delays of individual channels with the necessary attosecond precision. In the implementation presented here, rough optical path adjustment is possible for all optical channels along with fine adjustment (via piezoelectric stages) and is implemented for $Ch_{DUV}$, $Ch_{Vis-UV}$, and $Ch_{NIR}$. These elements enable field synthesis with attosecond resolution.

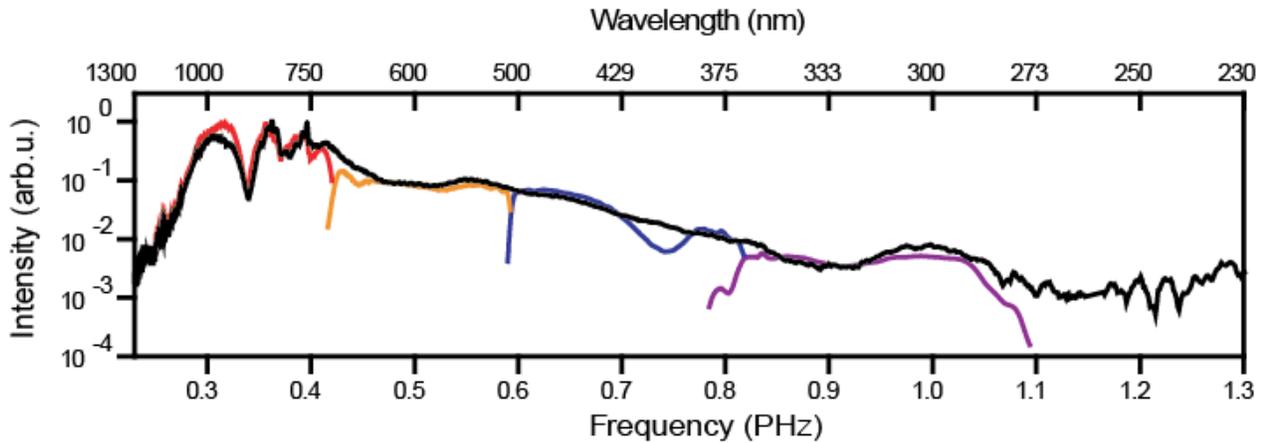

Figure 17. Supercontinuum spectrum after the HCF is shown as a black line, while the spectra of the individual channels, $Ch_{NIR}$, $Ch_{Vis}$, $Ch_{Vis-UV}$, and $Ch_{DUV}$ are shown as red, orange, blue, and violet lines, respectively.

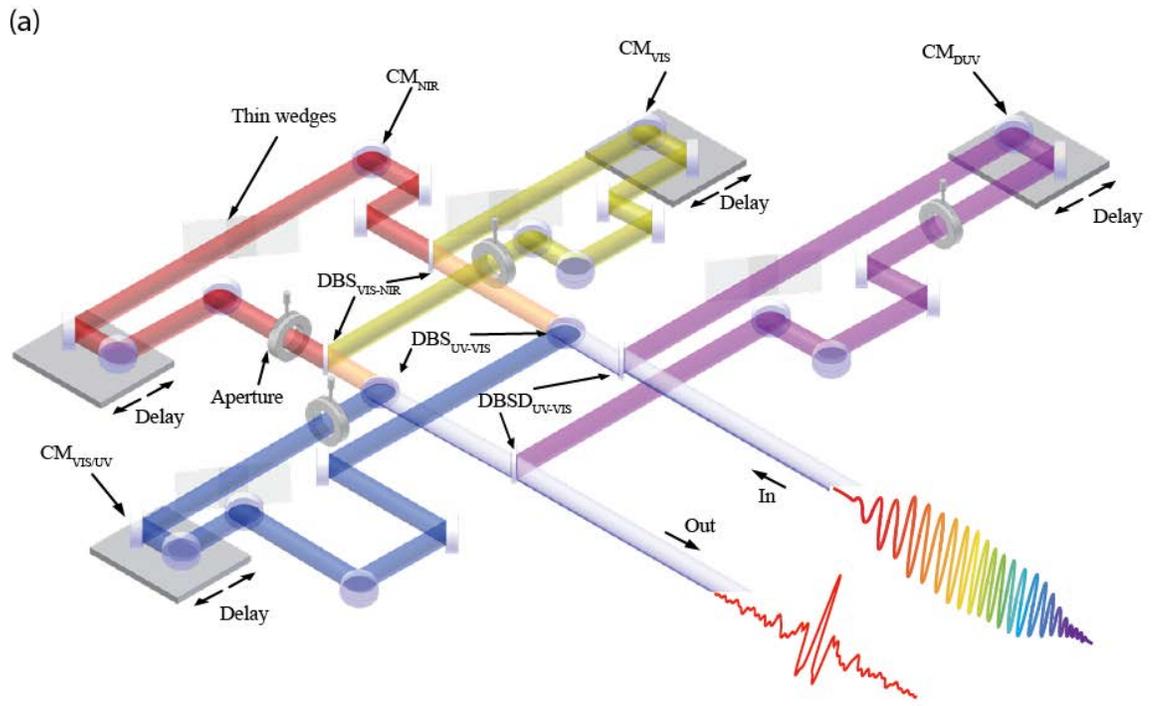

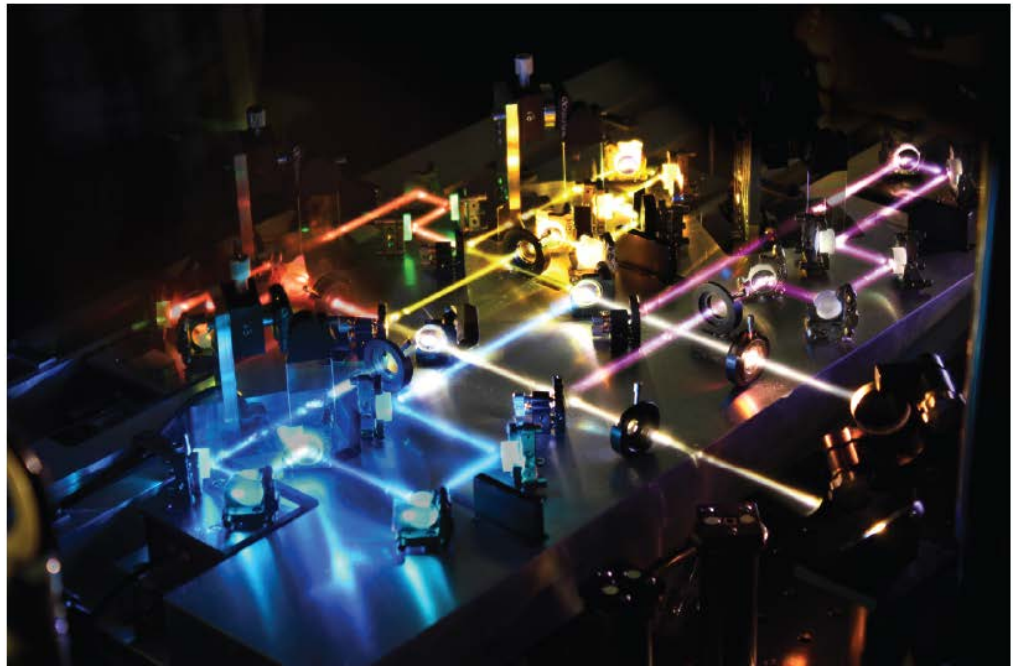

Figure 18: Attosecond light synthesizer (a) Schematic representation of a prototypical four-channel light field synthesizer. (b) Photograph (perspective) of the four-channel light field synthesizer in operation.

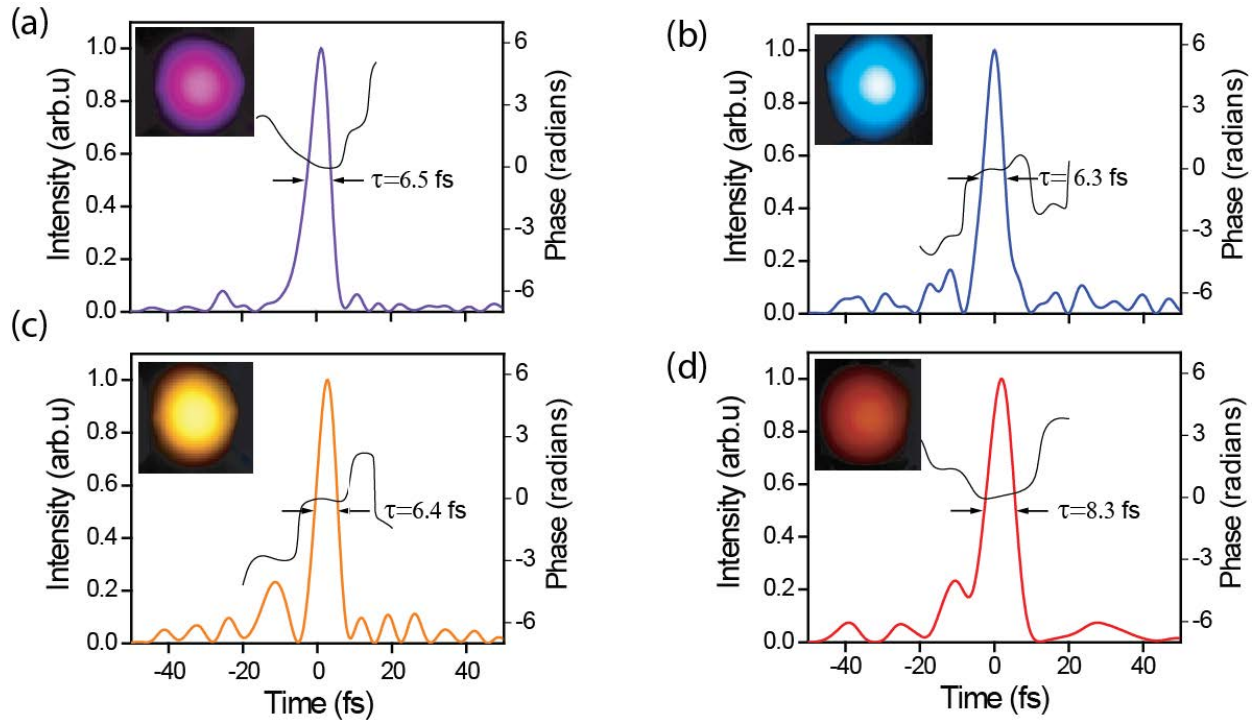

Figure 19: Temporal characterization for the pulses of the constituent channels in the four-channel synthesizer. The retrieved temporal profile for the constituent channels pulses (a) $Ch_{DUV}$, (b) $Ch_{Vis}$, (c) $Ch_{Vis-UV}$ and, and (d) $Ch_{NIR}$.

The generation of the OAP is not possible solely by superimposing the four channels at the exit of the synthesizer; the relative amplitudes of the different spectral bands need to be controlled to achieve a quasi-equal spectral intensity of all the channels. This can be attained by adjusting the aperture opening introduced in the path of each individual channel or by the introduction of band-pass filters.

Essentially, the four channels of the synthesizer are stabilized both passively and actively. The active phase-locking is introduced to the synthesizer to ensure the attosecond stability of the generated optical attosecond pulse, which has been proven experimentally [27]. Recently, we have demonstrated and characterized the optical attosecond pulse by attosecond streaking, which utilizes the generation of XUV attosecond electron pulses through High Harmonic Generation (HHG) [27]. In this case, adjusting the relative intensities between the channels actively inside the synthesizer was not possible since this would reduce the pulse intensity to below the threshold limit of the HHG process. To this end, the adjustment of the ultrabroadband spectrum intensity was done passively by an optical element, which we called metallic–dichroic–metallic mirror; for more details, see [27]. This mirror allowed for the reshaping of the supercontinuum spectrum to the desired relative intensities over two octaves leading to the generation of the optical attosecond pulse (Fig. 20).

The measured streaking spectrogram of the optical attosecond pulse is illustrated in Fig. 20. In the instantaneous intensity, which is shown in Fig. 20 (c), the central field crest is ~ 5 times more intense than those for the adjacent half cycles. The $\tau_{FWHM}$ duration of the central field crest is ~ 380 as. This optical attosecond pulse has a central wavelength of ~ 530 nm. We note that these optical attosecond pulses are generated reproducibly.

This optical attosecond pulse opens a new avenue in the field of attosecond physics. This pulse has been used to study and control the bound electron motion in atomic systems in real time [27]. Moreover, we plan to use this unique tool to generate an isolated attosecond electron pulse, as we will explain next.

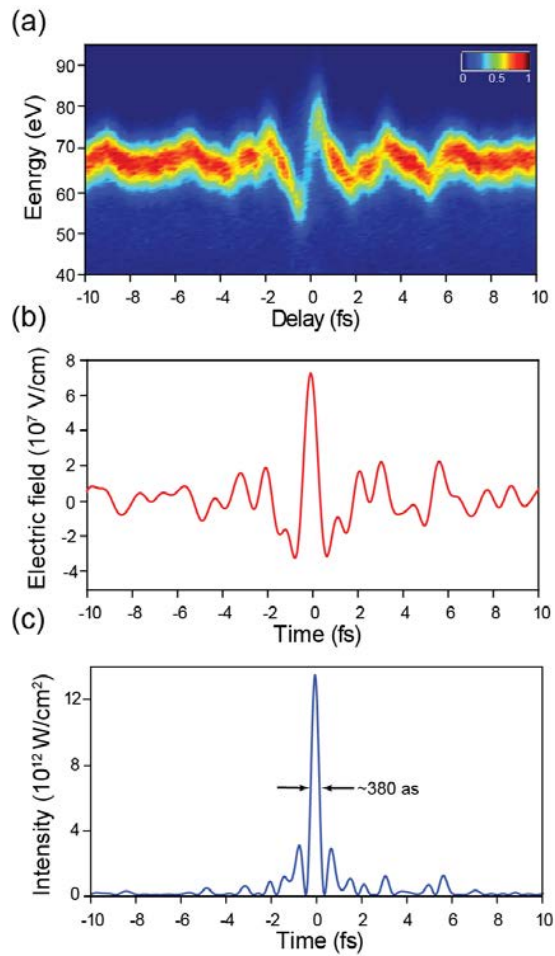

Figure 20: Optical Attosecond Pulse. (a) The sampled streaking spectrogram, (b) the retrieved electric field, and (c) the instantaneous intensity of the Optical Attosecond Pulse (OAP). The $\tau_{FWHM}$ duration of OAP is ~ 380 as.

*8.2. Isolated attosecond electron pulse*

As mentioned earlier, the duration of the gated electron pulse depends on the temporal gated window, which is determined by the optical gating pulse. Therefore, the optical attosecond pulse could provide the sub-femtosecond temporal gating resolution required to generate isolated attosecond electron pulses with the optical gating approach. Moreover, extremely high phase stability (< 1 fs) can be achieved between the pump and gating laser pulses by phase-locking, which was demonstrated experimentally in [27], to attain the attosecond temporal resolution in electron microscopy.

Hence, we perfomed a simple theoretical calculation, where a compensated 10 fs electron pulse (the generation of this short electron pulse was recently reported in [130]) is gated by a few-cycle optical pulse (10 fs), as well as by the optical attosecond pulses (380 as), as explained previously [27]. This was done by calculating the probability of a single electron emitting or absorbing photons through interaction with the surface of a nanostructure, following the steps outlined elsewhere [166]. The intensity of the optical pulse field is kept lower than the saturation threshold. In short, the electron–photon interaction is mediated by an evanescent electromagnetic field induced by the optical gating pulse hitting the surface. The spatial distribution of the evanescent field is determined by the optical properties and the geometry of the nanostructure. For simplicity, we assume an effective one-dimensional spatial distribution of the evanescent field.

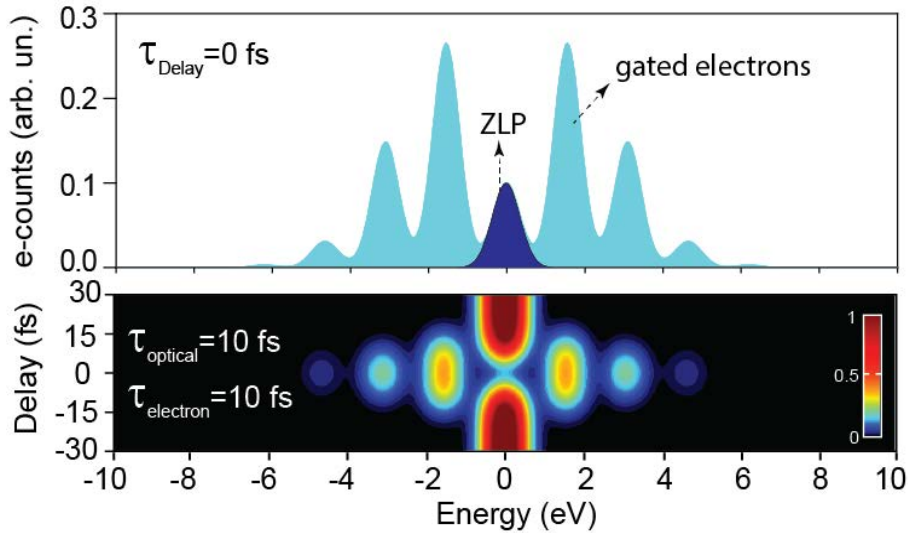

Figure 21. Few-cycle optical gating of electron pulse. The calculated gated electron spectrogram of the 10 fs electron pulse with the 10 fs visible laser pulse centred at 800 nm is shown in the bottom part, and the outline of the electron spectrum at a delay time of 0 fs is shown in the top part. The gated electron spectrum is shown in cyan and the ZLP spectrum is shown in blue.

Then, the strength of the evanescent field coupled with the electrons is given by [166]

$$E_z(z,t) = E_0 e^{-|z|/\xi} e^{-i\omega t - (t+\tau)^2/\tau_p^2}, \qquad (2)$$

where $E_0$ is the peak strength of the evanescent field, $\xi$ is the penetration depth into vacuum, $\omega$ is the photon frequency, $\tau$ is the delay between the optical and electron pulses, and $\tau_p$ is the duration of the optical pulse. The temporal evolution of the electrons in the presence of the evanescent field can be described by a time-dependent Schrödinger equation including a light–matter coupling Hamiltonian with a classical electromagnetic field, whose solution can be found self-consistently by solving the Lippmann–Schwinger equation (Eq. (4) in [166]). By expanding the electron wave function into momentum eigenstates corresponding to gaining or losing a certain number of photons, the Lippmann–Schwinger equation can be solved. The expansion coefficients are given in a recursive manner (Eq. (9) in [166]). With the expansion coefficients, the probabilities of electrons gaining/losing L photons can be computed from [166]

$$P_L = \sum_{N=|L|}^{\infty} \sum_{N`=|L|}^{\infty} C_L^N \left(C_L^{N`}\right)^* \frac{\exp\left\{-\frac{(N+N`)(\tau/\tau_p)^2}{1+[(N+N`)/2](\tau_e/\tau_p)^2}\right\}}{\sqrt{1+\left[\frac{(N+N`)}{2}\right](\tau_e/\tau_p)^2}}, \qquad (1)$$

where N and N` are the possible numbers of total scattering events, $C_L^N$ is the expansion coefficient of the electron wave function based on the momentum eigenstates corresponding to gaining or losing a certain number of photons, $\tau$ is the delay between the electron and "gating" optical pulses, $\tau_e$ is the electron pulse duration, and $\tau_p$ is the laser "gating" pulse duration.

The calculated electron energy spectrogram of the optical coupling between the 10 fs electron pulse and the 10 fs optical pulse (I = 3 × 10$^{12}$ W/m$^2$) is shown in Fig. 21. The 10 fs resolution is defined by the original electron pulse duration; the optical gating approach is necessary to achieve the extremely high phase stability between the pump (optical) and probe (gated electron) pulses for potential time-resolved electron microscopy experiments.

On the other hand, the attosecond optical gating of 10 fs electron pulses utilizing the OAP allows the generation of isolated attosecond electron pulses (Fig. 22). For such a broadband "gating" optical pulse, the gating medium (nanostructure) should have a broad frequency response (e.g., aluminum nanostructures, which support surface plasmon with lifetimes on the attosecond scale), and the gated electron spectral peaks are expected to exhibit energy broadening since the OAP spans more than two octaves (spectral FWHM = 1.75 eV), as shown in Figs. 22(a) and (b).

The optical gating efficiency in this case can be defined as the ratio of the number of gated electrons to the total number of electrons of the original electron pulse, which depends on the durations of both the original electron pulse and optical pulse.

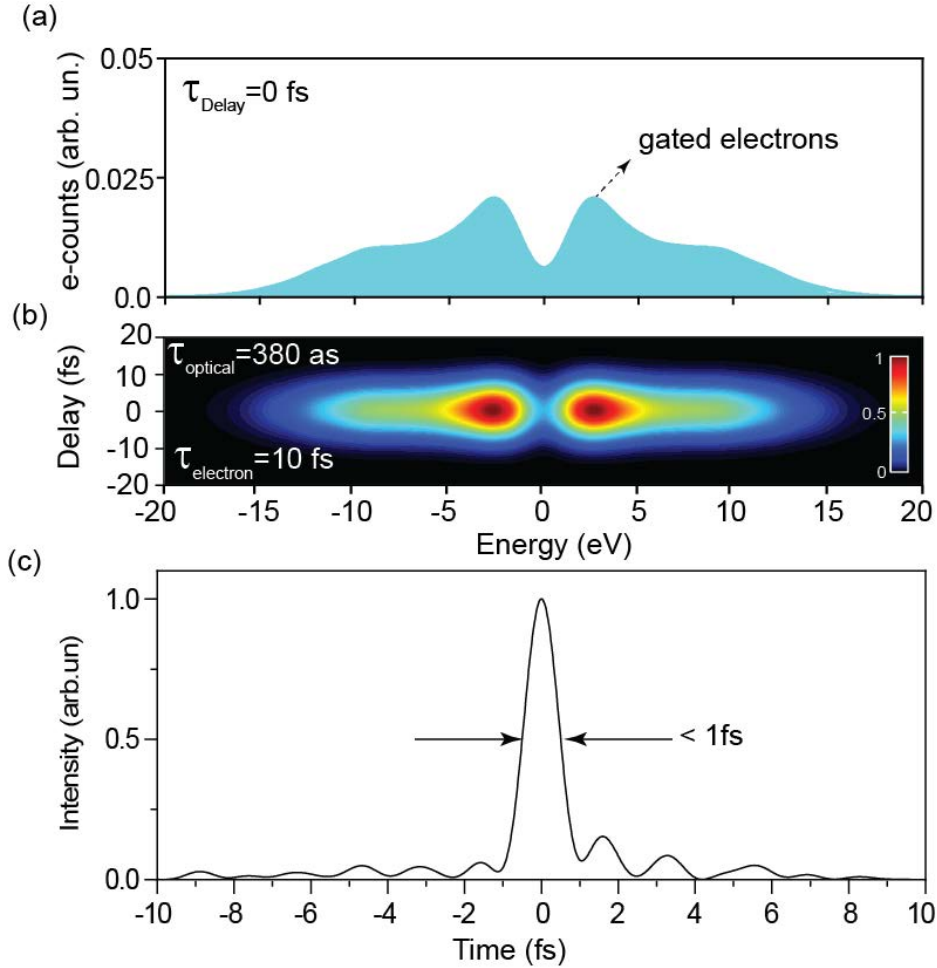

Figure 22. Attosecond optical gating of electron pulse. (a) Calculated spectrum of the gated electron of 10 fs electron pulse with OAP pulse centred at 530 nm at a delay time = 0 fs. (b) Calculated full spectrogram from -20 to 20 fs. The spectral width of the coupling peaks is broader than the coupling peaks in case of multicycle pulse (figure 21), since the OAP spectral FWHM=1.75 eV> the ZLP spectral FWHM= 1 eV. (c) Sub-femtosecond temporal window and the temporal profile of the OAP to generate the isolated attosecond electron pulses by optical gating.

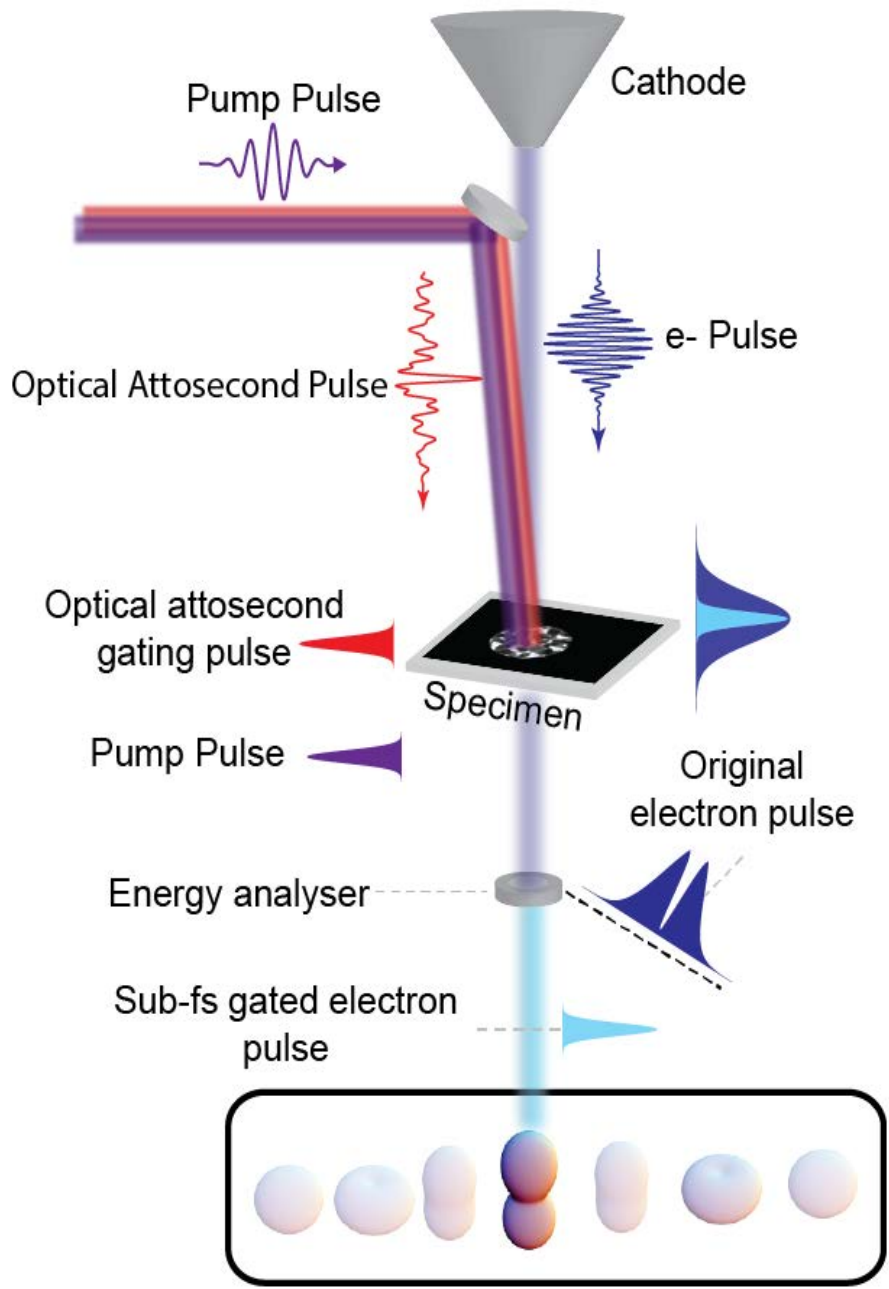

Fig. 23. Attomicroscopy: Imaging the electron motion in real time. Illustration of Attomicroscopy setup. The optical attosecond pulse is used to provide the sub-femtosecond optical gating window for the original electron pulse in order to generate an isolated attosecond electron pulse. The optical gating and the generation take place at the sample stage. The generated gated pulse can be filtered out later, providing a desired probe resolution to image the electron motion in real time.

For the potential time-resolved Attomicroscopy experiment (illustrated in Fig. 23), the isolated attosecond "gated" electron pulse on the sample stage can be filtered out and utilized as a "probe", while another optical pulse can be utilized as a "pump" to ultimately image the electron motion in real time. The isolated attosecond electron pulses can be generated and characterized right on the surface of the nanostructure sample, which allows for maintaining the attosecond resolution at the measured position (the dispersion of the gating optical pulse can be controlled by the light field synthesizer). In addition, a nanogrid of aluminum can be used as a nanostructure sample for attosecond optical gating of the electrons. This grid can be located directly above the sample under study. In this way, the generated attosecond electron pulse will not suffer from the temporal broadening due to the space-charge effect in the attosecond electron pulse generation. The attosecond resolution of the probe electron pulse can be preserved up to 1 cm propagation distances from the grid for electron pulses accelerated at 30 keV and assuming a single electron for a beam diameter of 50 μm. In either case, the nanometer spatial resolution is inherent in our generated attosecond pulse since the generation occurs on the nanometer grid.

Extremely high temporal resolution and stability can be achieved by the phase-locking of the laser "pump" pulse and the optical "gating" pulse. In addition, the phase-locking of the "gating" optical attosecond pulse is essential to minimize the timing jitter and the fluctuation of the "gated" electron pulse intensity. The achievement of attosecond–nanometer spatiotemporal resolution in electron microscopy will pave the way for establishing the new field of "Attomicroscopy". It has an enormous potential for femtosecond and attosecond imaging applications in different areas and could eventually enable the recording of a movie that shows electron motion in the act.

*8.3. Imaging electron motion in real time*

The imaging of electron motion utilizing the sub-cycle and attosecond electron pulse in solid state and atoms has been studied theoretically in detail elsewhere [29, 167]. Shao et al. [167] demonstrated the capabilities of imaging the breathing and wiggling modes of the electronic motion in the H atom by sub-femtosecond electron pulses. In this work, to explore the ability of electron motion imaging, the authors invoked the Robicheaux's general scattering theory of coherent matter beams [17] in ultrafast electron diffraction simulations [14, 15]. The key idea for describing the time-dependent scattering process is to use a coherent wave function consisting of wave packets of both the projectile electron and the target. By forming these wave packets, the projectile electron and the target are localized in space and in time, so the scattering events can be defined and analyzed properly. The authors separated the theoretical analysis of the scattering intensities, which carry the target information, into kinematical and dynamical aspects of the scattering. However, the authors defined two necessary conditions for the time-resolved electron diffraction measurements of the electron motion, which are: (i) the electron pulses should have a broader bandwidth compared to the energy scale of the electronic motion, and (ii) the electron pulse should not suffer from energy dispersion (it should have a smooth phase over the entire pulse bandwidth) so it can preserve its temporal resolution.

On the other hand, the Attomicroscopy technique can be used for imaging the electron motion dynamics indirectly. For example, the attosecond electron diffraction can be utilized to image the charge migration in molecules ionized by an optical pulse. This charge transfer is theoretically predicted to evolve on a scale of a few femtoseconds [168]. In this case, the change migration through the molecule will affect the bond length, which can be resolved in both time and spatial domains by time-resolved electron diffraction measurements. Furthermore, the imaging of these dynamics can determine the subsequent relaxation pathways of the molecule. The electron and structural dynamics of molecules in gas phase can be imaged by Attomicroscopy. However, to achieve the desired number of electrons in the gated attosecond electron pulse for a better signal-to-noise ratio, the initial electron pulse duration has to be on the order of several femtoseconds.

In conclusion, the generation of the attosecond electron pulses and the establishment of the Attomicroscopy will open new avenues and allow for a great number of femtosecond and attosecond electron imaging applications in different areas and could eventually enable the recording of a movie that shows electron motion in the act.

**Acknowledgment**

This work was supported by The University of Arizona and it is dedicated to Ahmed Zewail. I would like to thank T Karam and J S Baskin for their fruitful scientific discussions. I am grateful to the Gordon and Betty Moore Foundation for supporting my research activities at the Physical Biology Center for Ultrafast Science and Technology at Caltech.